\documentclass[twoside,11pt]{article}

\usepackage[accepted, specialissue]{melba}

\usepackage{amsmath,amsfonts}
\usepackage[acronym,toc,shortcuts]{glossaries}
\newacronym{ai}{AI}{artificial intelligence}
\newacronym{xai}{xAI}{explainable artificial intelligence}
\newacronym{cam}{CAM}{class activation map}
\newacronym{cnn}{CNN}{convolutional neural network}
\newacronym{ffn}{FFN}{feed-forward network}
\newacronym{opg}{OPG}{orthopantogram}
\newacronym{sota}{SOTA}{state-of-the-art}
\newacronym{gpu}{GPU}{graphics processing unit}
\newacronym{iou}{IoU}{Intersection over Union}
\newacronym{roar}{ROAR}{Remove and Retrain}

\usepackage[english, capitalise]{cleveref}
\usepackage{graphicx}
\usepackage{chngcntr}
\usepackage{caption}
\usepackage{subcaption}
\usepackage{float}
\usepackage{cite}
\usepackage{comment}
\usepackage{color,soul}
\usepackage{enumitem}

\usepackage{tabularx} % For width control
\usepackage{makecell} % For line breaks within cells
\usepackage{booktabs} % For better-looking horizontal rules
\newcommand{\eg}{\textit{e}.\textit{g}.}
\newcommand{\ie}{\textit{i}.\textit{e}.}

\melbaid{2024:023}  % This is provided upon by the publishing editor
\doi{https://doi.org/10.59275/j.melba.2024-ebd3}
\melbaauthors{Rheude et al.}  % Note: this one is also used to set the pdf 'authors' metadata
\volume{2}
\firstpageno{2089} % Communicated by the publishing editor
\melbayear{2024}  % The publication year
\datesubmitted{03/2024}  % Date submitted to MELBA: mm/yyyy
\datepublished{10/2024}  % Today's date: mm/yyyy

\melbaspecialissue{Interpretability of Machine Intelligence in Medical Image Computing \\(iMIMIC) 2023}
\melbaspecialissueeditors{Mauricio Reyes, Jaime Cardoso, Jayashree Kalpathy-Cramer, Nguyen Le Minh, Pedro Abreu, José Amorim, Wilson Silva, Mara Graziani, Amith Kamath}

\ShortHeadings{Leveraging CAM Algorithms for Explaining Medical Semantic Segmentation}{Rheude et al.}

\title{Leveraging CAM Algorithms for Explaining Medical Semantic Segmentation}

\author{\firstname Tillmann \surname Rheude \\  % start right after \author{, or there will be an extra space
	\addr TU Darmstadt, Darmstadt, Germany \\
Fraunhofer Institute for Computer Graphics Research (IGD), Darmstadt, Germany
	\AND
\firstname Andreas \surname Wirtz \\  % start right after \author{, or there will be an extra space
	\addr Fraunhofer Institute for Computer Graphics Research (IGD), Darmstadt, Germany
	\AND
\firstname Arjan \surname Kuijper \\  % start right after \author{, or there will be an extra space
	\addr TU Darmstadt, Darmstadt, Germany  \\
Fraunhofer Institute for Computer Graphics Research (IGD), Darmstadt, Germany
	\AND
\firstname Stefan \surname Wesarg \\  % start right after \author{, or there will be an extra space
	\addr Fraunhofer Institute for Computer Graphics Research (IGD), Darmstadt, Germany
}

\begin{document}

% top matter
\maketitle

% abstract
\begin{abstract}%   <- trailing '%' for backward compatibility of .sty file
    \Glspl{cnn} achieve prevailing results in segmentation tasks nowadays and represent the state-of-the-art for image-based analysis. However, the understanding of the accurate decision-making process of a \gls{cnn} is rather unknown. The research area of \gls{xai} primarily revolves around understanding and interpreting this black-box behavior. One way of interpreting a \gls{cnn} is the use of \glspl{cam} that represent heatmaps to indicate the importance of image areas for the prediction of the \gls{cnn}. For classification tasks, a variety of \gls{cam} algorithms exist. But for segmentation tasks, only two \gls{cam} algorithms for the interpretation of the output of a \gls{cnn} exist. We propose a transfer between existing classification- and segmentation-based methods for more detailed, explainable, and consistent results which show salient pixels in semantic segmentation tasks. The resulting \textit{Seg-HiRes-Grad CAM} is an extension of the segmentation-based \textit{Seg-Grad CAM} with the transfer to the classification-based \textit{HiRes CAM}. Our method improves the previously-mentioned existing segmentation-based method by adjusting it to recently published classification-based methods. Especially for medical image segmentation, this transfer solves existing explainability disadvantages. The code is available at ~\url{https://github.com/TillmannRheude/SegHiResGrad_CAM}.
\end{abstract}

\glsreset{cnn}
\glsreset{xai}
\glsreset{cam}

% keywords
\begin{keywords}
	Deep Learning, Explainable Artificial Intelligence, Gradient-Based Methods, Medical Image Segmentation
\end{keywords}

% Introduction (or first section)
\section{Introduction}
%\noindent \hl{TODO: melbaid, doi, firstpage, nodatesubmitted, datepublished, melbaspecialissue, melbaspecialissueeditors} \\

Even if the (mathematical) theory of neural networks' training process is well known, the exact reasoning behind why neural networks derive at a particular prediction is rather hard to interpret \citep{selvaraju_grad-cam_2017}. The black box behavior of neural networks complicates the understanding of their decision-making process. Especially for medical tasks, understanding this process is essential \citep{Chen_Gomez_Huang_Unberath_2022}. Here, the question would be, \eg, where does the prediction that there is a tumor in the image come from? Ideally, the prediction comes from image regions containing the tumor. However, it would also be possible that the prediction is triggered by other factors that do not contribute to the presence or absence of the tumor. 

Algorithms for \gls{xai} can be classified in different ways. Local explanation algorithms derive a reasoning of individual predictions $f(x)$ of a model $f$ in contrast to global explanation algorithms which derive a reasoning with only the model $f$ and without the need of any predictions $f(x)$ of the model \citep{agarwal_2021}. Model-specific algorithms are only applicable to certain models \citep{agarwal_2021}. On the other hand, model-agnostic algorithms like SHAP \citep{lundberg_shap} are applicable regardless of the model choice \citep{agarwal_2021}. Gradient-based methods use the gradients of the neural network as a proxy in comparison to pertubation-based methods which derive the explanation by altering the input data \citep{ivanovs_2021}. In summary, algorithms which use \glspl{cam}\footnote{For a particular algorithm, we use the italic font while it is not italicized when we refer to the general idea of class activation maps (CAMs).} are therefore classified as local because of the use of individual input datapoints on which a heatmap is placed and they are model-specific since they cannot be applied to any model like a simple regression model. It has to be noted that \gls{cam}-based algorithms do not have to be gradient-based but they are most of the times nowadays as explained in \cref{related_work}. With our work, we want to improve the \gls{sota} for explainable, local, model-specific and gradient-based algorithms for image segmentation via \glspl{cam}.

To generate a heatmap of an input image that interprets the model output, two well-known methods were proposed in the past: saliency maps \citep{zeiler_visualizing_2014} and \glspl{cam} \citep{zhou_learning_2016}. 
A saliency map, \eg, created by Saliency Mapping \citep{simonyan_deep_2014}, DeconvNets \citep{zeiler_visualizing_2014} or Guided Backpropagation \citep{springenberg_guided_backprop}, is a heatmap for showing the importance of individual image pixels. These maps are based on different calculcations of the gradients regarding the input features.
\glspl{cam} are also heatmaps but they are based on the activations of the last or one specific convolutional layer of a \gls{cnn}. They are able to visualize the spatial positioning and expansion and are used as a weighting factor for the activations of the feature maps. In summary, both methods produce similar heatmaps for interpretation, but they differ in how they are calculated. While saliency maps are classified as input-level approaches, \gls{cam} algorithms are classified as output-level approaches \citep{draelos_use_2021}. The advantage of \gls{cam} algorithms is that they do not need the propagation through all layers and that they are more robust in contrast to saliency maps \citep{draelos_use_2021}. Robustness is especially important in domains where the precise localization and visualization of influential image regions play a pivotal role, such as medical imaging applications.

\begin{figure}[ht]
    \centering
    
    \includegraphics[trim={0 0 0 2.2cm},clip,width=\linewidth]{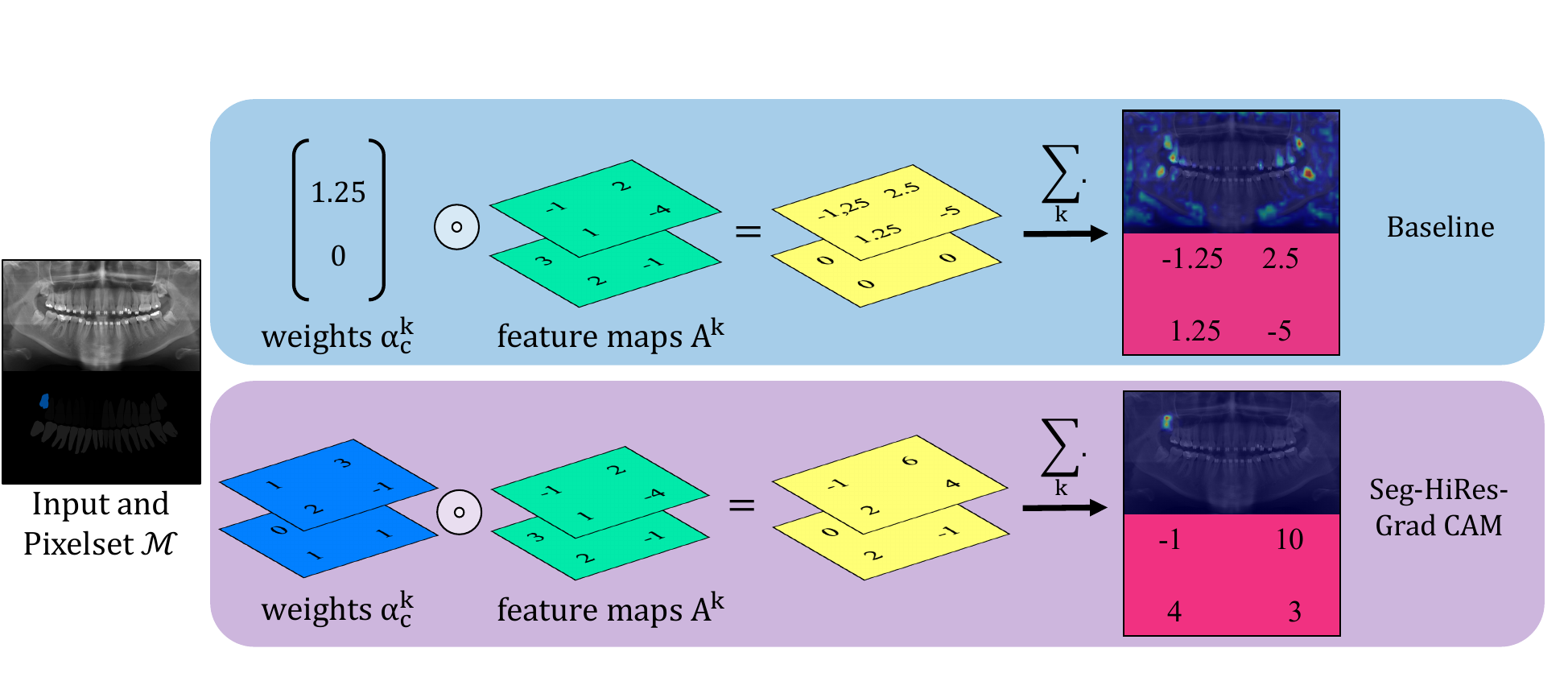} % {figures/flowchart_new.png} % 

    \caption{Calculation flow of \textit{Seg-Grad CAM} \citep{vinogradova_towards_2020} (\textbf{top row}) and our proposed \textit{Seg-HiRes-Grad CAM} (\textbf{bottom row}) based on the computations and flowchart of \textit{HiRes CAM} \citep{draelos_use_2021} with $K$ feature maps, dimensions $D_1$ and $D_2$ of the feature maps and the average weight values $\alpha_c^k$. Gradients (blue) are multiplied with activation maps (green). The sum (red) of the product (yellow) is upscaled for the final heatmap before ReLU and after ReLU (above red square). On the left, the respective input image and semantic segmentation of the U-Net \citep{ronneberger_u-net_2015} is shown. It is striking that the output \glspl{cam} are very different. ReLU is not visualized to ensure a better comparability.}
    \label{fig:flowchart_seggradcam_vs_seghiresgradcam}
\end{figure}

Algorithms for the visualization of \glspl{cam} are dominated by classification-based work (\cref{tab:cam_publications}). For segmentation purposes, a transfer is needed and already proposed with \textit{Seg-Grad CAM} \citep{vinogradova_towards_2020}. Nevertheless, this transfer is based on \textit{Grad CAM} \citep{selvaraju_grad-cam_2017} and for classification tasks \textit{Grad CAM} has the disadvantage of certain inaccuracies \citep{draelos_use_2021}. These inaccuracies include, \eg, the visualization of regions that are not used by the \gls{cnn} for the prediction \citep{draelos_use_2021}. \\ 
We propose \textit{Seg-HiRes-Grad CAM} to solve the before-mentioned inaccuracies in the segmentation case, too. \textit{Seg-HiRes-Grad CAM} is based on \textit{Seg-Grad CAM} and transfers the classification-based \textit{HiRes CAM} \citep{draelos_use_2021} to segmentation tasks (\cref{fig:flowchart_seggradcam_vs_seghiresgradcam}). 

% Reviewer Comment Addition
The paper is structured as follows: We start by stating related work in \cref{related_work} for being able to understand the method derivation in \cref{method}. Afterwards, we show qualitative results in \cref{results} and discuss possible caveats in \cref{discussion}. At the end, we draw a conclusion and point out possible future work in \cref{conclusion}.

\section{Related Work}
\label{related_work}
The first algorithm of visualizing \glspl{cam} after the before-mentioned methods for saliency maps is called \textit{CAM} \citep{zhou_learning_2016}. The idea of \textit{CAM} is the use of the weights of the last \gls{ffn} in a \gls{cnn} as weightings for the feature maps of the last convolutional layer. 
For a more complex calculation and more interpretable visual results, \textit{Grad CAM} \citep{selvaraju_grad-cam_2017} is proposed which complements the weight calculation with the average of the respective gradients of a neural network. But, \textit{Grad CAM} still visualizes certain inaccuracies in some cases \citep{draelos_use_2021}. One of the most present inaccuracies is that there are cases in which \textit{Grad CAM} visualizes regions the \gls{cnn} did not actually use \citep{draelos_use_2021}. 
\textit{HiRes CAM} \citep{draelos_use_2021} solves these inaccuracies. In \textit{HiRes CAM}, the gradients are not averaged anymore as in \textit{Grad CAM}. \textit{HiRes CAM} uses the raw gradients for calculating the weights, which are multiplied with the activations of the \gls{cnn}. The before-mentioned algorithms represent the most important publications for visualizing \glspl{cam}. Further publications which are related to the topic of visualizing \glspl{cam} but which are not necessary to understand our proposed approach are listed in \cref{tab:cam_publications} (left column). 

The before-mentioned methods refer to the visualization of \glspl{cam} in classification tasks but not to the visualization in segmentation tasks. For the latter, \textit{Seg-Grad CAM}  is the first method for visualizing \glspl{cam}. However, the inaccuracies that occur in \textit{Grad CAM}, which were tackled by \textit{HiRes CAM}, are also present in \textit{Seg-Grad CAM}, since \textit{Seg-Grad CAM} is a modification to \textit{Grad CAM}. Consequently, we propose the transfer of the classification-based \textit{HiRes CAM} and the segmentation-based \textit{Seg-Grad CAM} resulting in \textit{Seg-HiRes-Grad CAM} (\cref{fig:flowchart_seggradcam_vs_seghiresgradcam}). The simultaneously published method \textit{Seg-XRes-CAM} \citep{Hasany_Petitjean_Meriaudeau_2023} is also based on the combination of \textit{Seg-Grad CAM} and \textit{HiRes CAM}. However, \textit{Seg-XRes-CAM} differs from \textit{Seg-HiRes-Grad CAM} as it includes a pooling layer in the calculation. This results in additional hyperparameters such as a window size. An evaluation for medical images and an ablation study regarding the pooling layer are not provided for \textit{Seg-XRes-CAM}.

\begin{table}
    \caption{Excerpt of popular algorithms for \glspl{cam} in classification tasks and the respective publications for segmentation tasks. Our proposed algorithm (bold) represents the transfer of \textit{HiRes CAM} \citep{draelos_use_2021} to segmentation tasks.}
    \centering 
    \begin{tabularx}{\textwidth}{|X|X|}
    \toprule
    \thead{Publications \\ for classification tasks} & \thead{Respective publications \\ for segmentation tasks} \\ 
    \hline
    \makecell{CAM \\ \citep{zhou_learning_2016}} & \makecell{\textit{n/a}} \\
    \makecell{Grad CAM \\ \citep{selvaraju_grad-cam_2017}} & \makecell{Seg-Grad CAM \\ \citep{vinogradova_towards_2020}} \\
    \makecell{Grad CAM++ \\ \citep{chattopadhay_grad-cam_2018}} & \makecell{\textit{n/a}} \\
    \makecell{XGrad CAM \\ \citep{fu_axiom-based_2020}} & \makecell{\textit{n/a}} \\
    \makecell{Ablation CAM \\ \citep{desai_ablation-cam_2020}} & \makecell{\textit{n/a}} \\
    \makecell{Score CAM \\ \citep{wang_score-cam_2020}} & \makecell{\textit{n/a}} \\
    \makecell{Eigen CAM \\ \citep{muhammad_eigen-cam_2020}} & \makecell{\textit{n/a}} \\
    \makecell{Layer CAM \\ \citep{jiang_layercam_2021}} & \makecell{\textit{n/a}} \\
    \makecell{FullGrad CAM \\ \citep{srinivas_full-gradient_2019}} & \makecell{\textit{n/a}} \\
    \makecell{LIFT CAM \\ \citep{jung_towards_2021}} & \makecell{\textit{n/a}} \\
    \makecell{HiRes CAM \\ \citep{draelos_use_2021}} & 
    \makecell{Seg-XRes-CAM \citep{Hasany_Petitjean_Meriaudeau_2023} \\ \textbf{Seg-HiRes-Grad CAM} (ours)} \\
    \bottomrule
    \end{tabularx}
    \label{tab:cam_publications}
\end{table}

\section{Method}
\label{method}

With our work, we want to improve the \gls{sota} of \gls{cam} algorithms for (medical) image segmentation. As elaborated in the previous \lcnamecref{related_work}, \glspl{cam} can be interpreted as heatmaps which can be placed upon the input image illustrating the areas which are more critical for the decision-making process. Since most of the work is done for classification tasks (\cref{tab:cam_publications}), we start empirically by explaining the mathematical theory behind these methods first and draw the transfer to segmentation-based algorithms afterwards. \textit{CAM} is described mathematically as follows with the heatmap $L^c_{CAM}$, weights $\alpha$ of the \gls{ffn}, the respective class $c$, feature maps $A$ and the respective number of the channel $k$ in the feature map \citep{simonyan_deep_2014}: 

\begin{equation}
    L^c_{CAM} = \sum_k \alpha^k_c A^k.
    \label{eq:cam}
\end{equation}

\noindent Improving the visual explainability, \textit{Grad CAM} \citep{selvaraju_grad-cam_2017} is proposed based on \textit{CAM}. The weighting ($\alpha$ in \cref{eq:cam}) is calculated with the respective gradients instead of only the weights of the \gls{ffn}. Consequently, the mathematical description changes as follows with the heatmap $L^c_{Grad CAM}$ and the ReLU function: 

\begin{equation}
    L^c_{Grad CAM} = \text{ReLU} \left( \sum_k \alpha_c^k A^k \right), 
    \label{eq:grad_cam}
\end{equation}

\noindent including the weights $\alpha$ with the number of pixels $N$, individual pixels $u,v$, and outputs $y^c$:

\begin{equation}
    \alpha_c^k = \frac{1}{N} \sum_{u, v} \frac{\partial y^c}{\partial A^k_{uv}}.
    \label{eq:grad_cam_weights}
\end{equation}

\noindent But recently, it was shown  that \textit{Grad CAM} visualizes regions in the image which do not contribute to the outcoming prediction \citep{draelos_use_2021}. Thus, \eg, in the classification task of the atelectasis (collapsed lung), \textit{Grad CAM} suggests that the neural network uses pixels which are located at the heart \citep{draelos_use_2021}. This is not explainable since the \gls{cnn} should look at the lung instead. To solve this inaccuracy, \textit{HiRes CAM} \citep{draelos_use_2021} is proposed which correctly visualizes the lung region in this example. The difference between these two \gls{cam} visualization methods is the mean calculation for the weights $\alpha$ (\cref{eq:grad_cam_weights}) \citep{draelos_use_2021}. It is proposed to calculate the weights with the same notation (\cref{eq:grad_cam_weights}) as follows instead \citep{draelos_use_2021}: 

\begin{equation}
    \alpha_c^k = \frac{\partial y^c}{\partial A^k}.
    \label{eq:hires_cam}
\end{equation}

\noindent So far, all equations refer to classification-based \glspl{cnn}. For segmentation-based \glspl{cnn}, a transfer is needed: In segmentation tasks, there is a label for every pixel and not only for every picture, as in classification tasks. As a consequence, Vinogradova et al. \citep{vinogradova_towards_2020} propose to modify $y^c$ in \cref{eq:grad_cam} and \cref{eq:grad_cam_weights} as follows with $\mathcal{M}$, the \texttt{\char`\"}set of pixel indices of interest in the output mask\texttt{\char`\"} \citep{vinogradova_towards_2020} and the respective, individual pixels $i, j$: 

\begin{equation}
    y^{c, new} = \sum_{i,j \in \mathcal{M}} y^c_{i, j}.
    \label{eq:seg_grad_cam_yc}
\end{equation}

\noindent With this modification, the pixel set $\mathcal{M}$ can be chosen in a flexible way and the classification-based \textit{Grad CAM} is transferred to the segmentation-based \textit{Seg-Grad CAM} \citep{vinogradova_towards_2020}. For instance, $\mathcal{M}$ can define all pixels of the image or only a certain amount of pixels up to only one certain pixel \citep{vinogradova_towards_2020}. On the other hand, this method still suffers from the inaccuracies of the original proposed \textit{Grad CAM} revealed by \textit{HiRes CAM} for classification purposes. To address this limitation of \textit{Seg-Grad CAM}, we propose \textit{Seg-HiRes-Grad CAM} by combining the segmentation-based \textit{Seg-Grad CAM} and the classification-based \textit{HiRes CAM} (\cref{fig:flowchart_seggradcam_vs_seghiresgradcam}). The weights $\alpha_c^k$ are not calculated by using the mean anymore, but the formula for the weights still involves the modification with the pixel set $\mathcal{M}$. This results in the following weight calculation formula: 

\begin{equation}
    \alpha_c^k = \frac{y^{c, new}}{\partial A^k} = \frac{\sum_{i,j \in \mathcal{M}} y^c_{i, j}}{\partial A^k}.
    \label{eq:seg_hires_grad_cam}
\end{equation}

\noindent For the segmentation task, every layer with feature maps (\ie, $A^k$) of a \gls{cnn} can be used, but it is common to use the deepest layer which contains the highest number of feature maps \citep{vinogradova_towards_2020}. The deepest layer represents most of the feature information, while higher layers represent more edge-like structures \citep{vinogradova_towards_2020}. Since our method is affected by the model performance, we have attached corresponding details and training results (even for further datasets) in \cref{tab:training_details} for improved transparency. We use commonly known, multi-class and semantic segmentation datasets such as Cityscapes \citep{cordts_cityscapes_2016}, Kits23 \citep{Heller_Isensee_et_al_2020} and OPG \citep{jader_deep_2018} to prove the method in medical and non-medical settings (\cref{tab:training_details}). These heterogeneous datasets contain varying amounts of data, classes, and also challenges such as homogeneous gray levels (OPG) or fine-grained details (Cityscapes, Kits23).

\section{Results}
\label{results}
\begin{figure}[ht]
    \centering
    \begin{subfigure}{0.32\textwidth}
      \includegraphics[width=\linewidth]{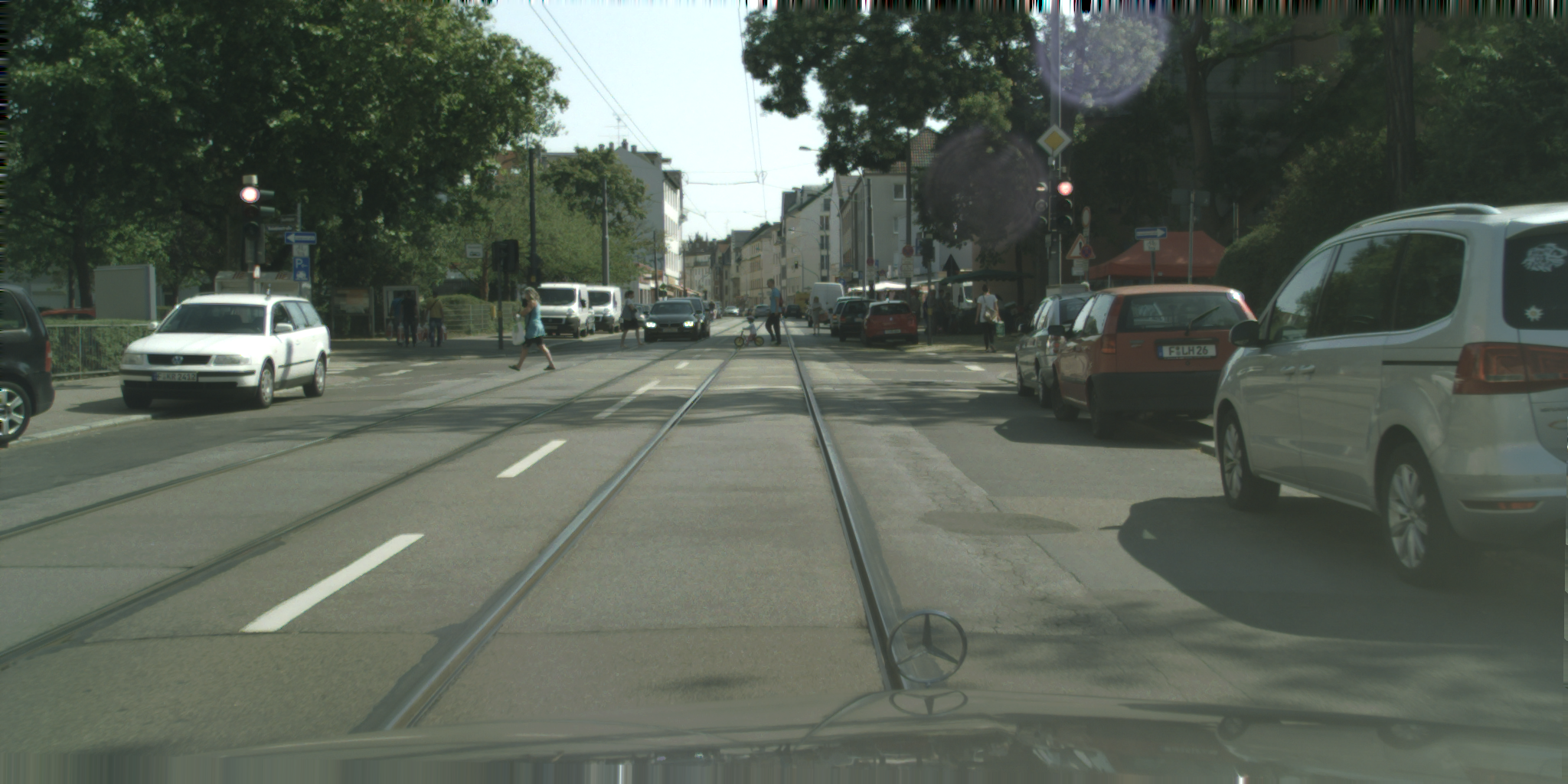}
      \caption{Input image}
    \end{subfigure}
    \begin{subfigure}{0.32\textwidth}
      \includegraphics[width=\linewidth]{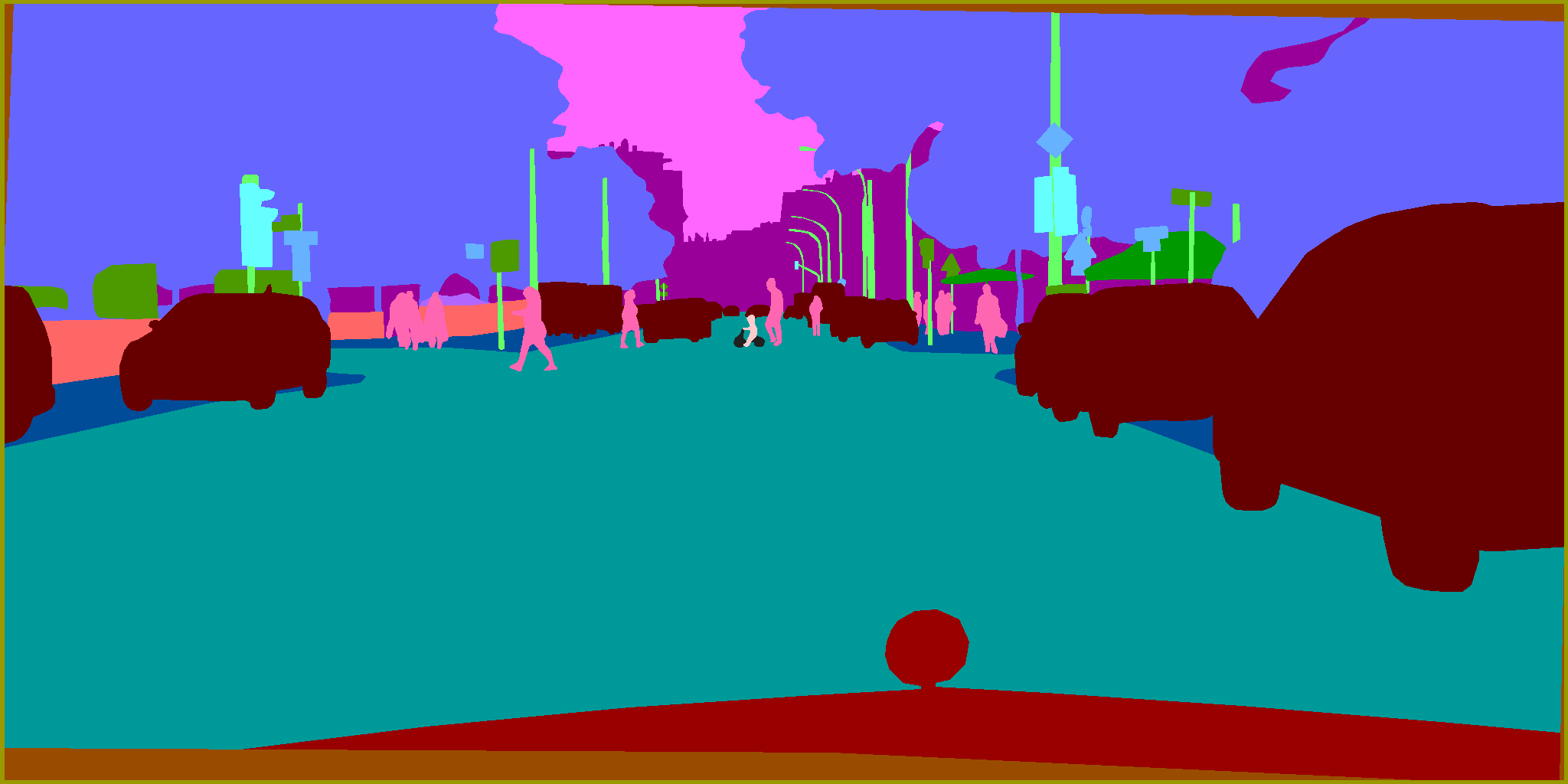}
      \caption{Ground truth}
    \end{subfigure}
    \begin{subfigure}{0.32\textwidth}
      \includegraphics[width=\linewidth]{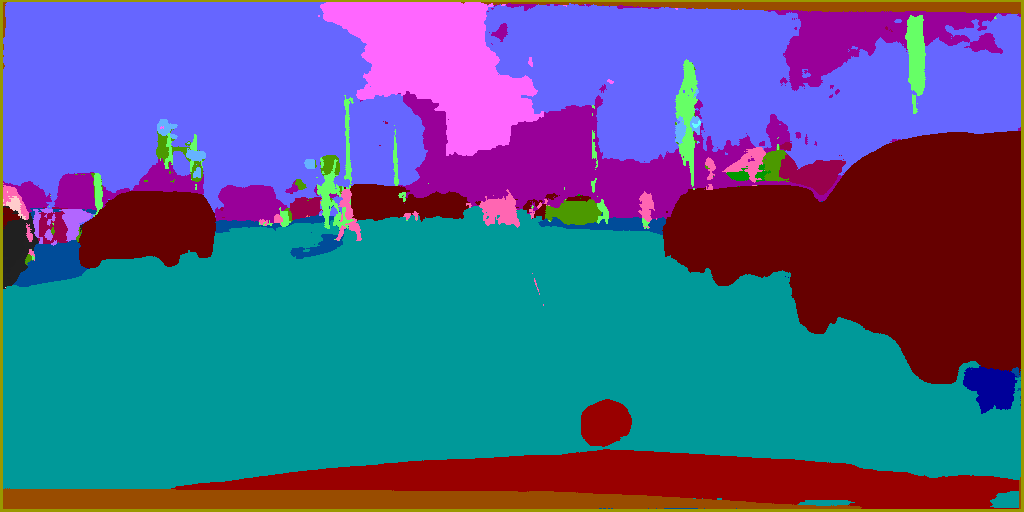}
      \caption{Prediction}
    \end{subfigure}
    
    \medskip
    
    \begin{subfigure}{0.32\textwidth}
      \includegraphics[width=\linewidth]{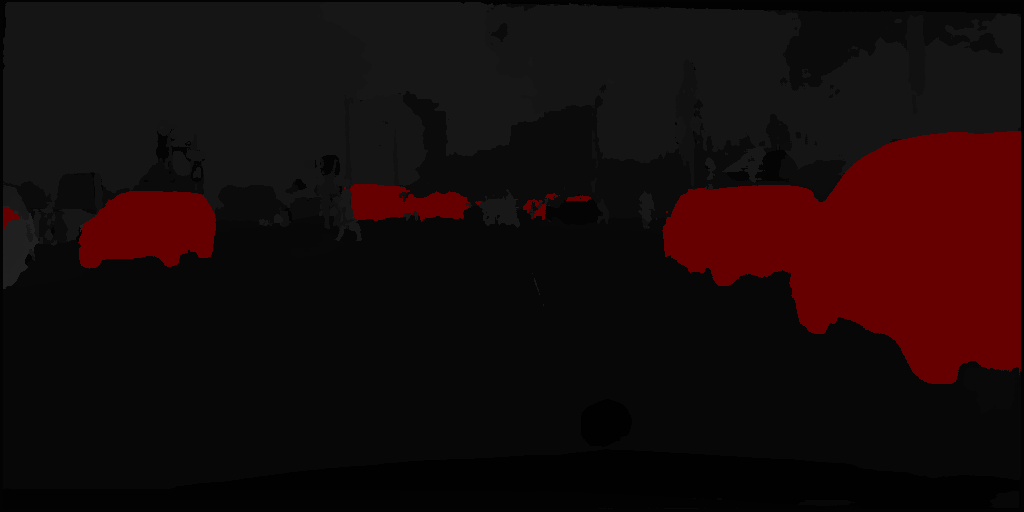}
      \caption{Pixel set $\mathcal{M}$}
    \end{subfigure}
    \begin{subfigure}{0.32\textwidth}
      \includegraphics[width=\linewidth]{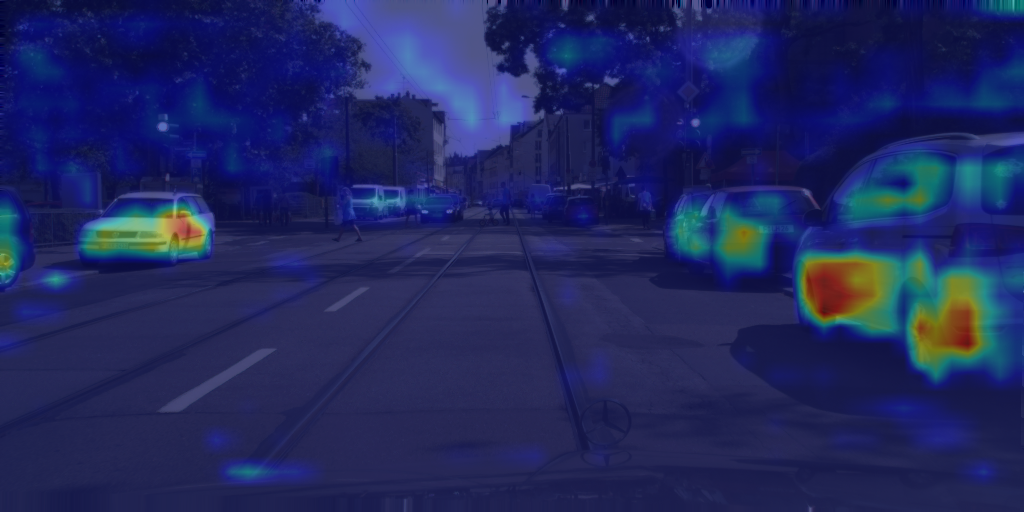}
      \caption{Seg-Grad CAM}
    \end{subfigure}
    \begin{subfigure}{0.32\textwidth}
      \includegraphics[width=\linewidth]{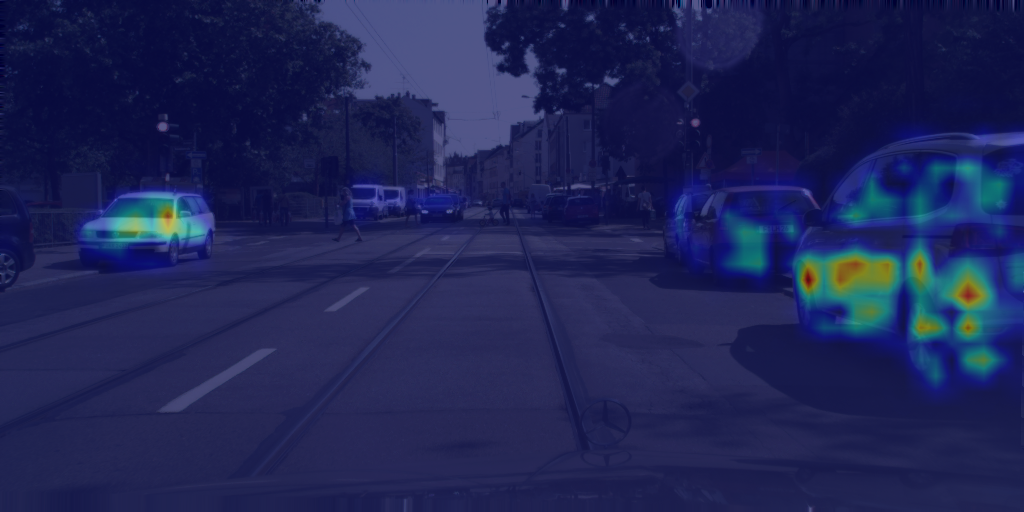}
      \caption{Seg-HiRes-Grad CAM}
    \end{subfigure}
    
    \caption{Comparison between \textit{Seg-Grad CAM} \citep{vinogradova_towards_2020} \textbf{(e)} and \textit{Seg-HiRes-Grad CAM} \textbf{(f)}. In this case, $\mathcal{M}$ equals the respective pixels of the prediction \textbf{(c)} for the car class \textbf{(d)}, which is similar to the ground truth \textbf{(b)}. The input image \textbf{(a)} from the Cityscapes dataset \citep{cordts_cityscapes_2016} is used since Vinogradova et al. \citep{vinogradova_towards_2020} use it.}
    \label{fig:cam_vs_hirescam_cityscapes}
\end{figure}

Comparing our proposed \textit{Seg-HiRes-Grad CAM} with the \textit{Seg-Grad CAM} baseline, the resulting heatmaps differ regarding their accuracy and explainability. In \cref{fig:cam_vs_hirescam_cityscapes}, the qualitative comparison with the Cityscapes dataset \citep{cordts_cityscapes_2016} for the car-class is illustrated. 

\begin{figure}[ht]
    \centering
    \begin{subfigure}{0.32\textwidth}
      \includegraphics[width=\linewidth]{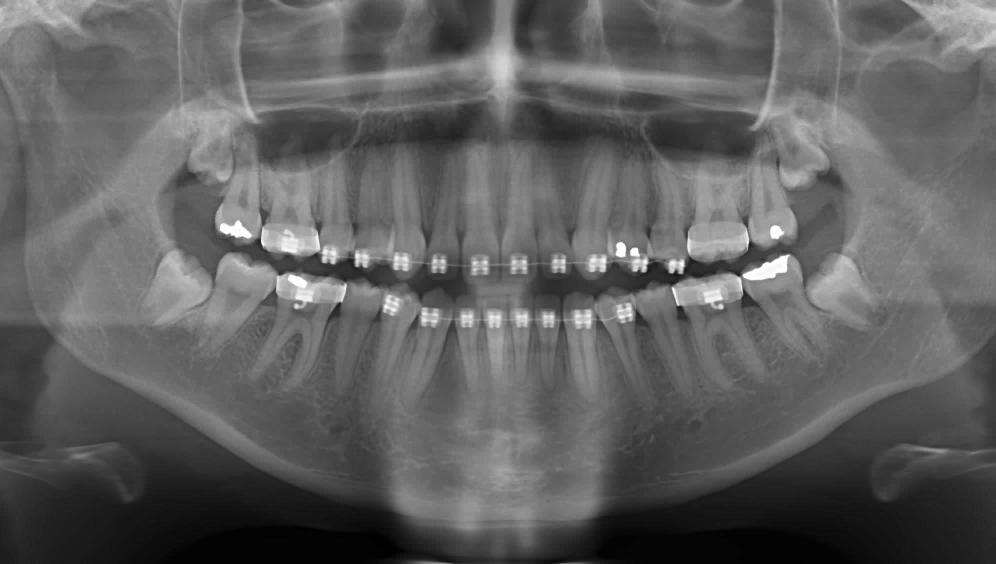}
      \caption{Input image}
    \end{subfigure}
    \begin{subfigure}{0.32\textwidth}
      \includegraphics[width=\linewidth]{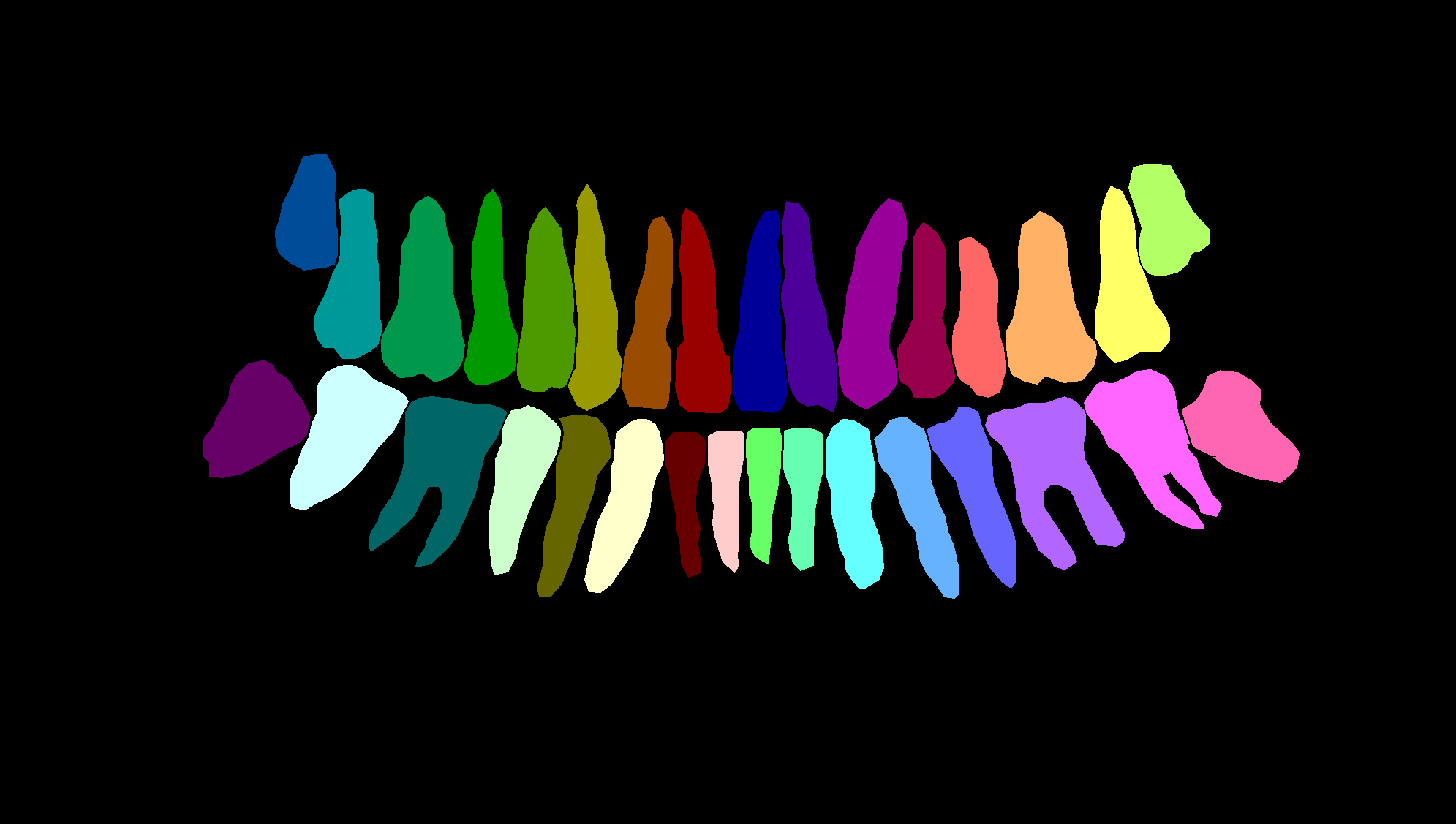}
      \caption{Ground truth}
    \end{subfigure}
    \begin{subfigure}{0.32\textwidth}
      \includegraphics[width=\linewidth]{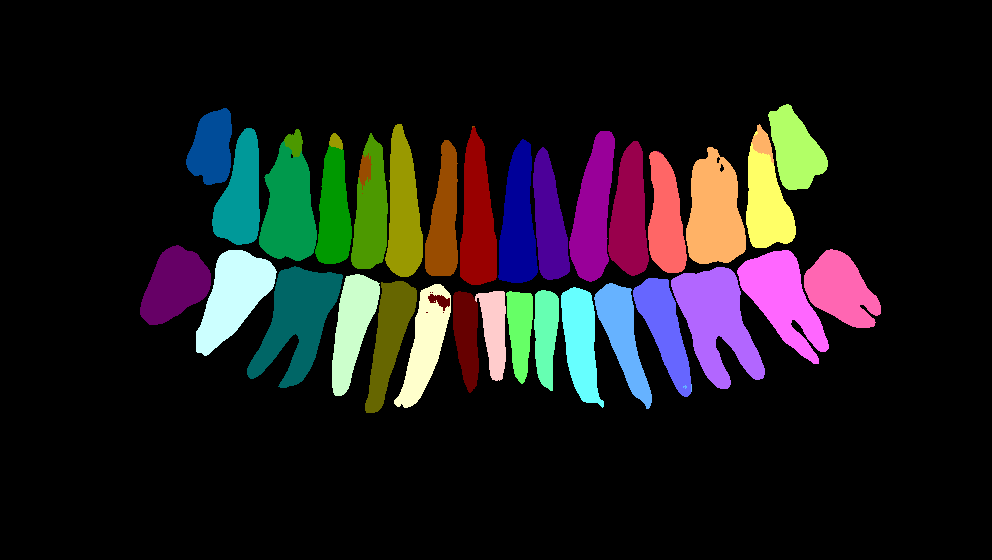}
      \caption{Prediction}
    \end{subfigure}
    
    \medskip
    
    \begin{subfigure}{0.32\textwidth}
      \includegraphics[width=\linewidth]{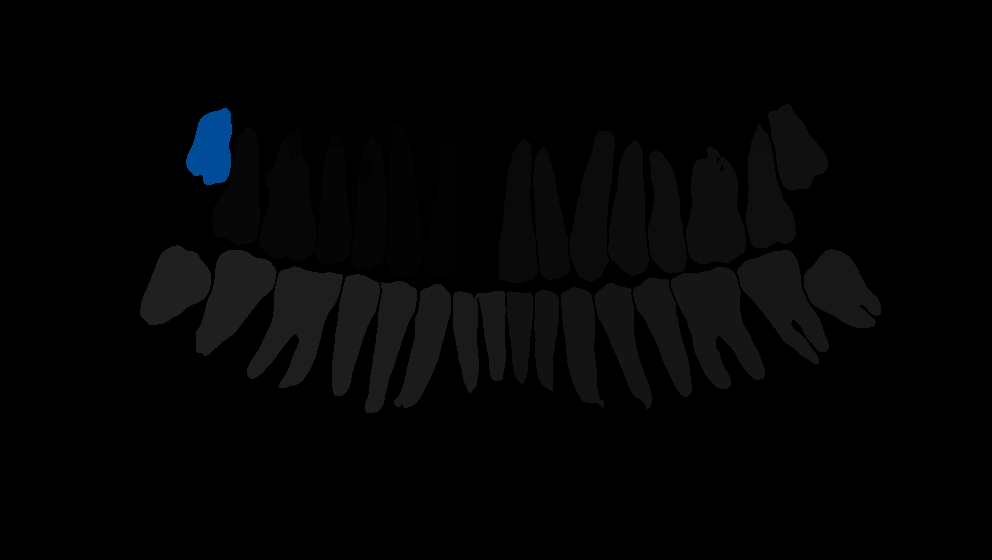}
      \caption{Pixel set $\mathcal{M}$}
    \end{subfigure}
    \begin{subfigure}{0.32\textwidth}
      \includegraphics[width=\linewidth]{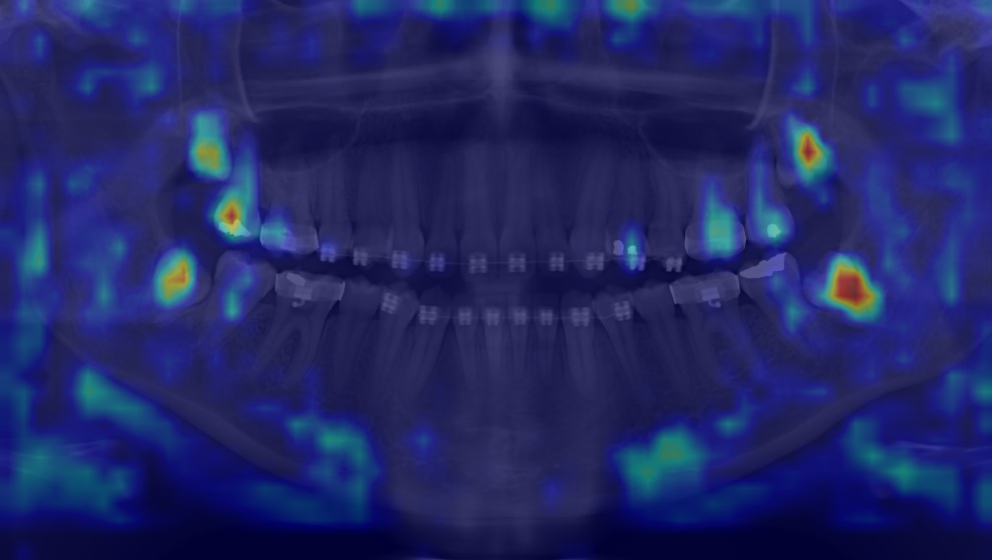}
      \caption{Seg-Grad CAM}
    \end{subfigure}
    \begin{subfigure}{0.32\textwidth}
      \includegraphics[width=\linewidth]{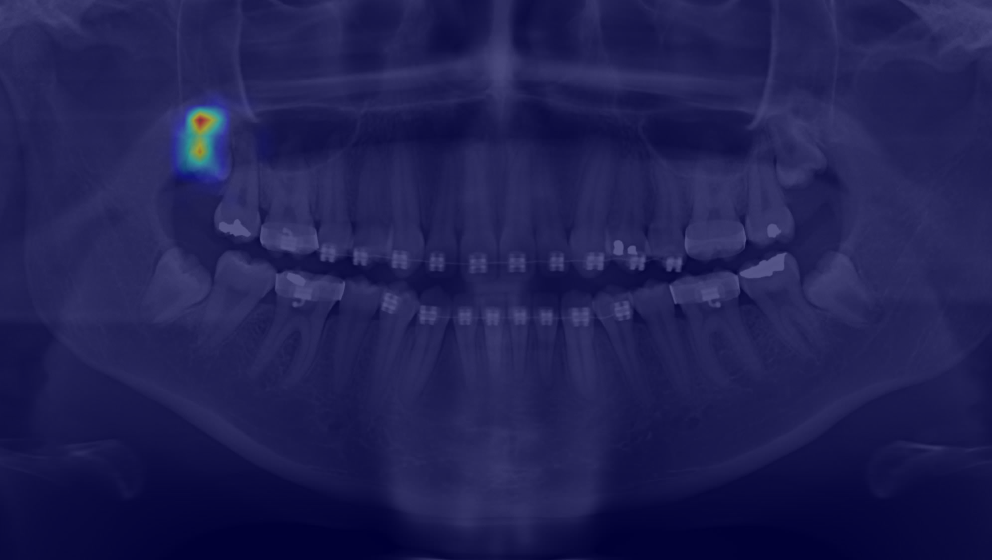}
      \caption{Seg-HiRes-Grad CAM}
    \end{subfigure}
    
    \caption{Comparison between \textit{Seg-Grad CAM} \citep{vinogradova_towards_2020} \textbf{(e)} and \textit{Seg-HiRes-Grad CAM} \textbf{(f)} for the upper right wisdom tooth (blue segmentation) \textbf{(d)}. In this case, $\mathcal{M}$ equals the respective pixels of the prediction \textbf{(c, d)}, which is similar to the ground truth \textbf{(b)}. The input image \textbf{(a)} comes from the \gls{opg} dataset \citep{jader_deep_2018}.}
    \label{fig:cam_vs_hirescam_opg_tooth_18}
\end{figure}

\begin{figure}[ht]
    \centering
    \begin{subfigure}{0.22\textwidth}
      \includegraphics[width=\linewidth]{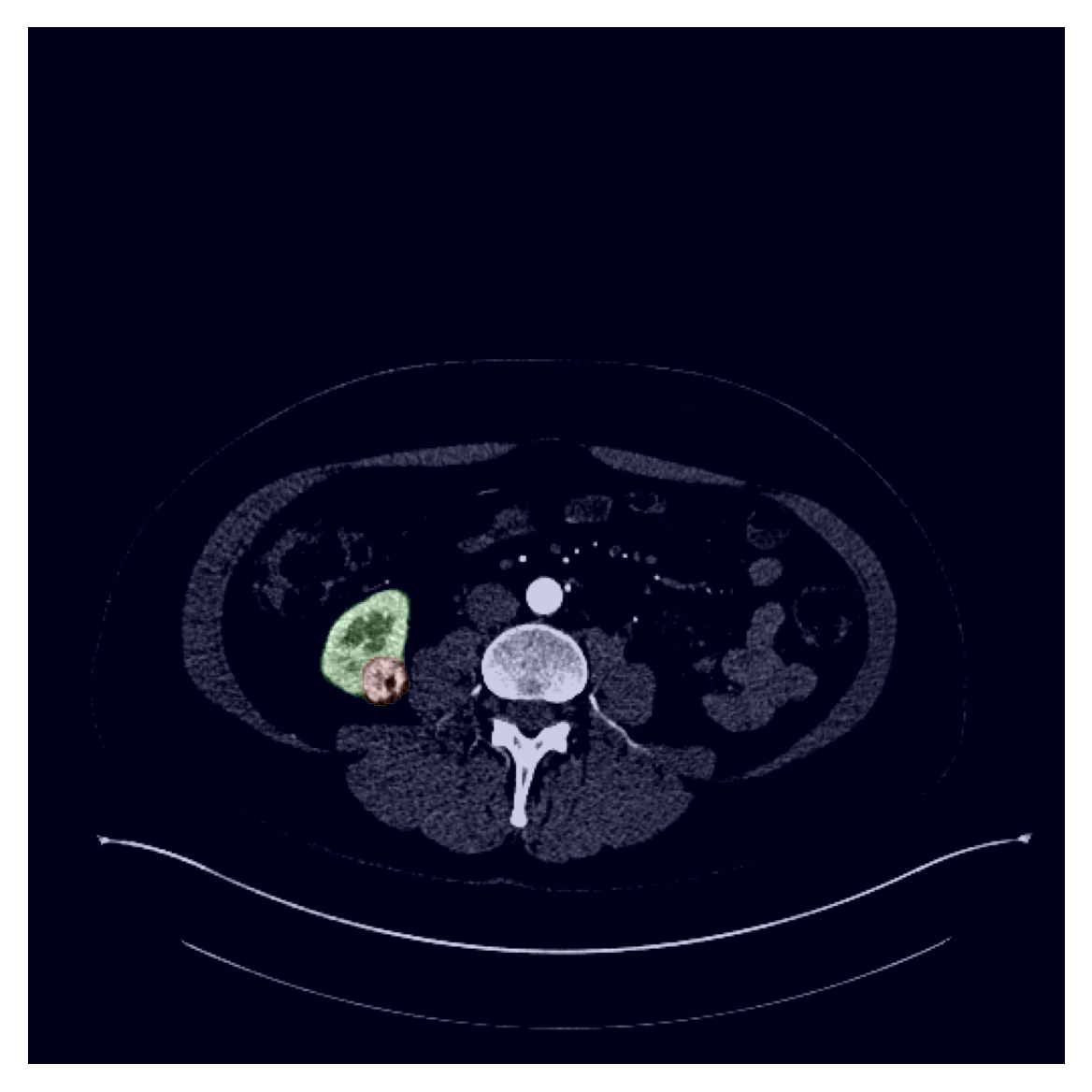}
      \caption{GT}
    \end{subfigure}
    \begin{subfigure}{0.22\textwidth}
      \includegraphics[width=\linewidth]{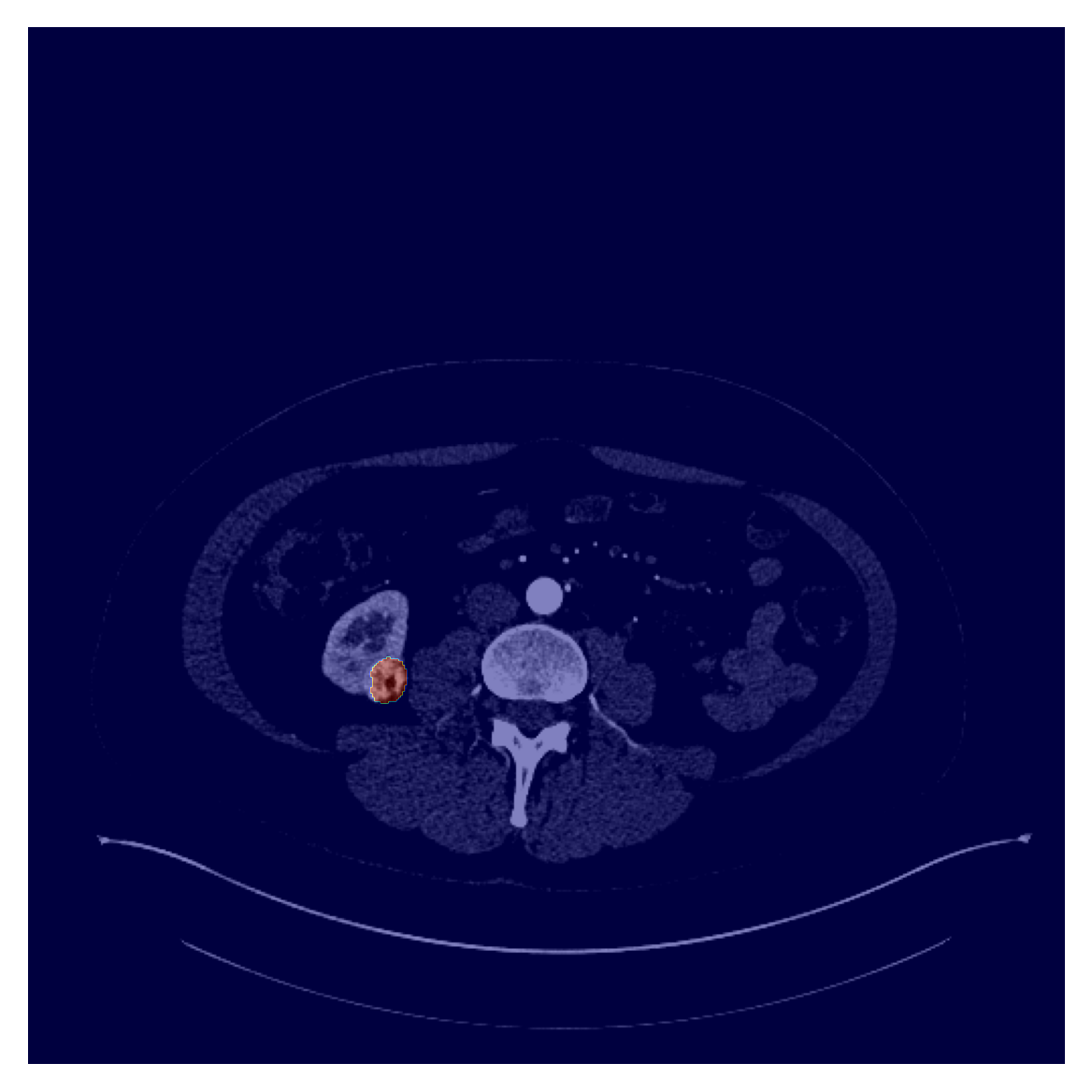}
      \caption{$\mathcal{M}$}
    \end{subfigure}    
    \begin{subfigure}{0.22\textwidth}
      \includegraphics[width=\linewidth]{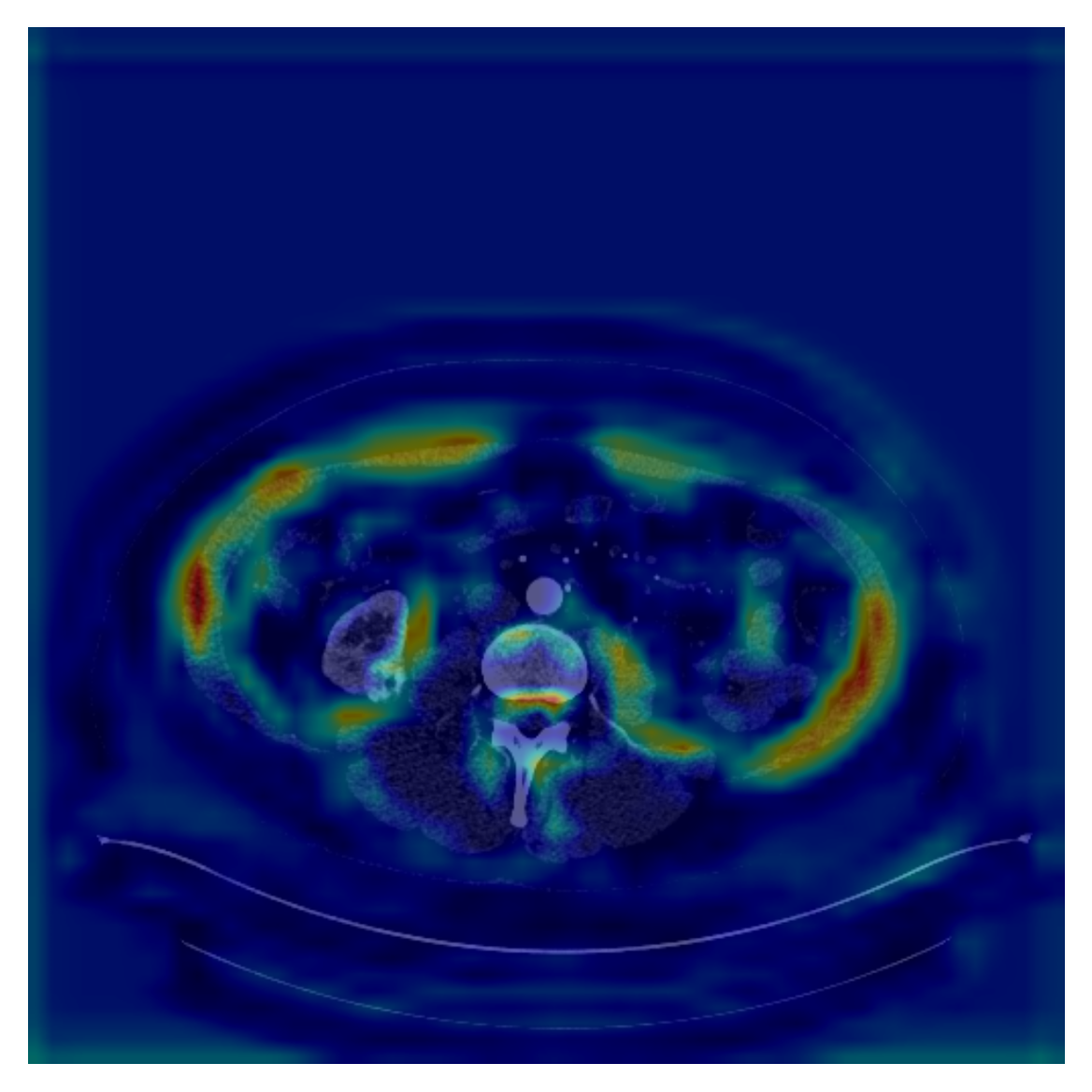}
      \caption{SGC}
    \end{subfigure}    
    \begin{subfigure}{0.22\textwidth}
      \includegraphics[width=\linewidth]{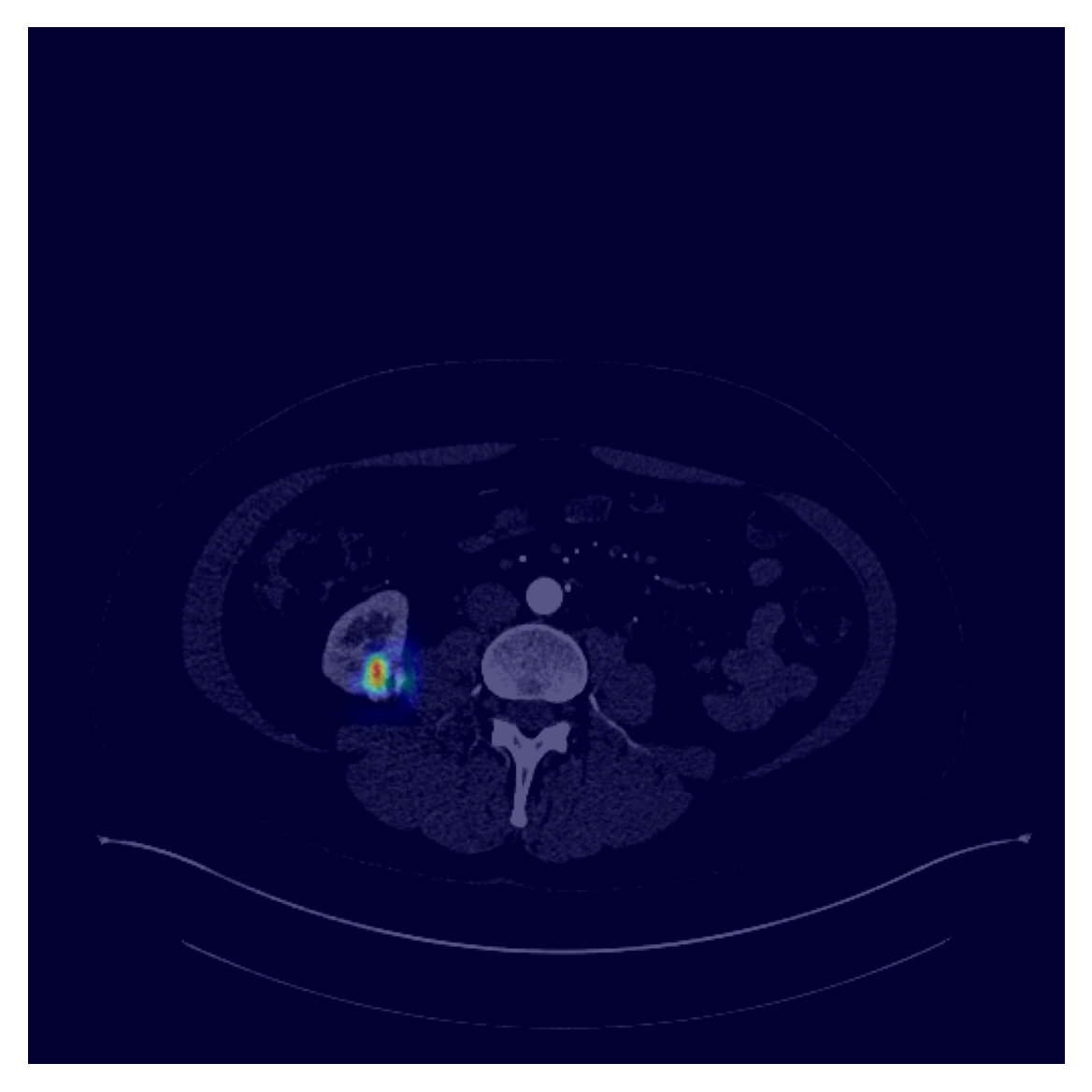}
      \caption{SHRGC}
    \end{subfigure}

    \begin{subfigure}{0.22\textwidth}
      \includegraphics[width=\linewidth]{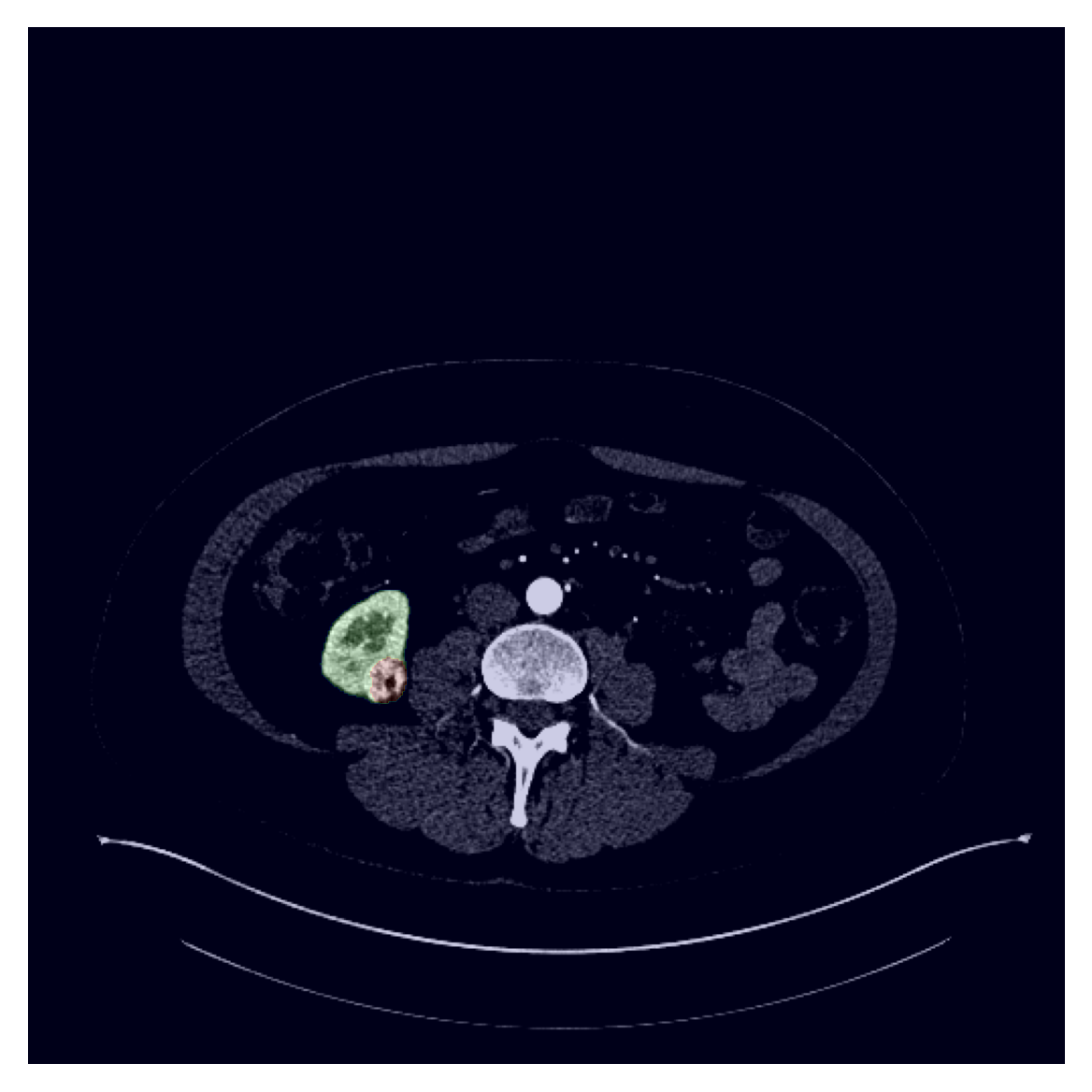}
      \caption{Pred}
    \end{subfigure}
    \begin{subfigure}{0.22\textwidth}
      \includegraphics[width=\linewidth]{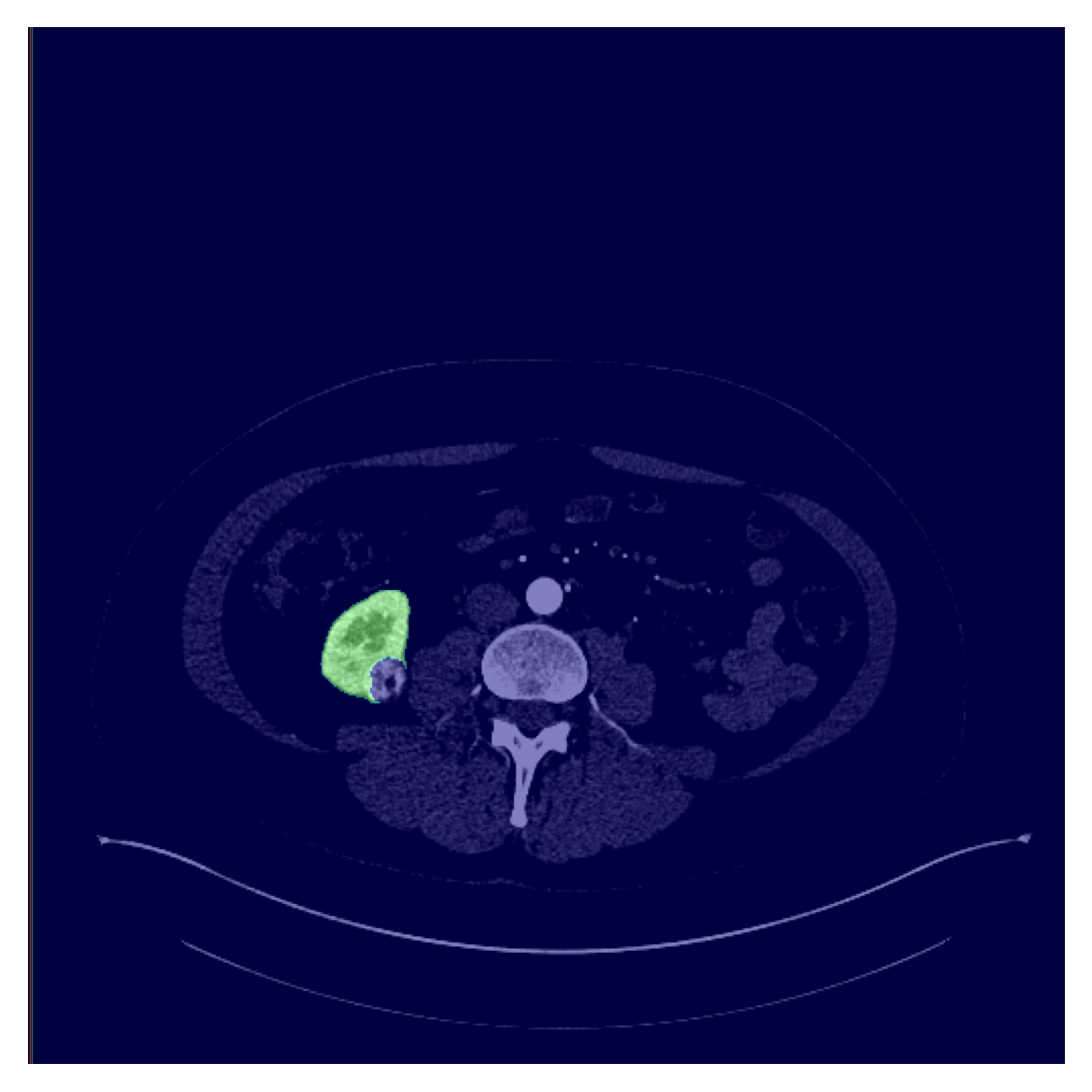}
      \caption{$\mathcal{M}$}
    \end{subfigure}
    \begin{subfigure}{0.22\textwidth}
      \includegraphics[width=\linewidth]{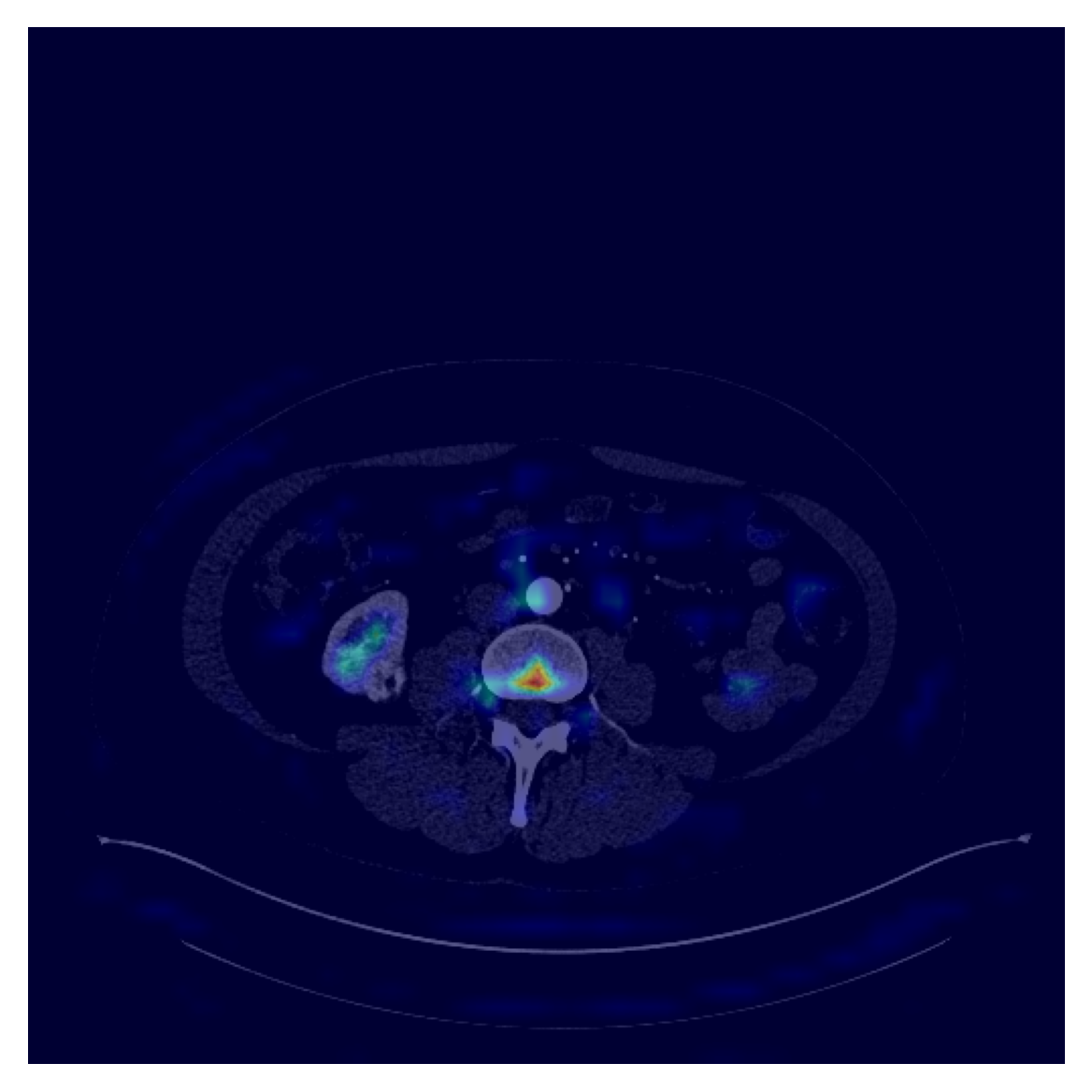}
      \caption{SGC}
    \end{subfigure}
    \begin{subfigure}{0.22\textwidth}
      \includegraphics[width=\linewidth]{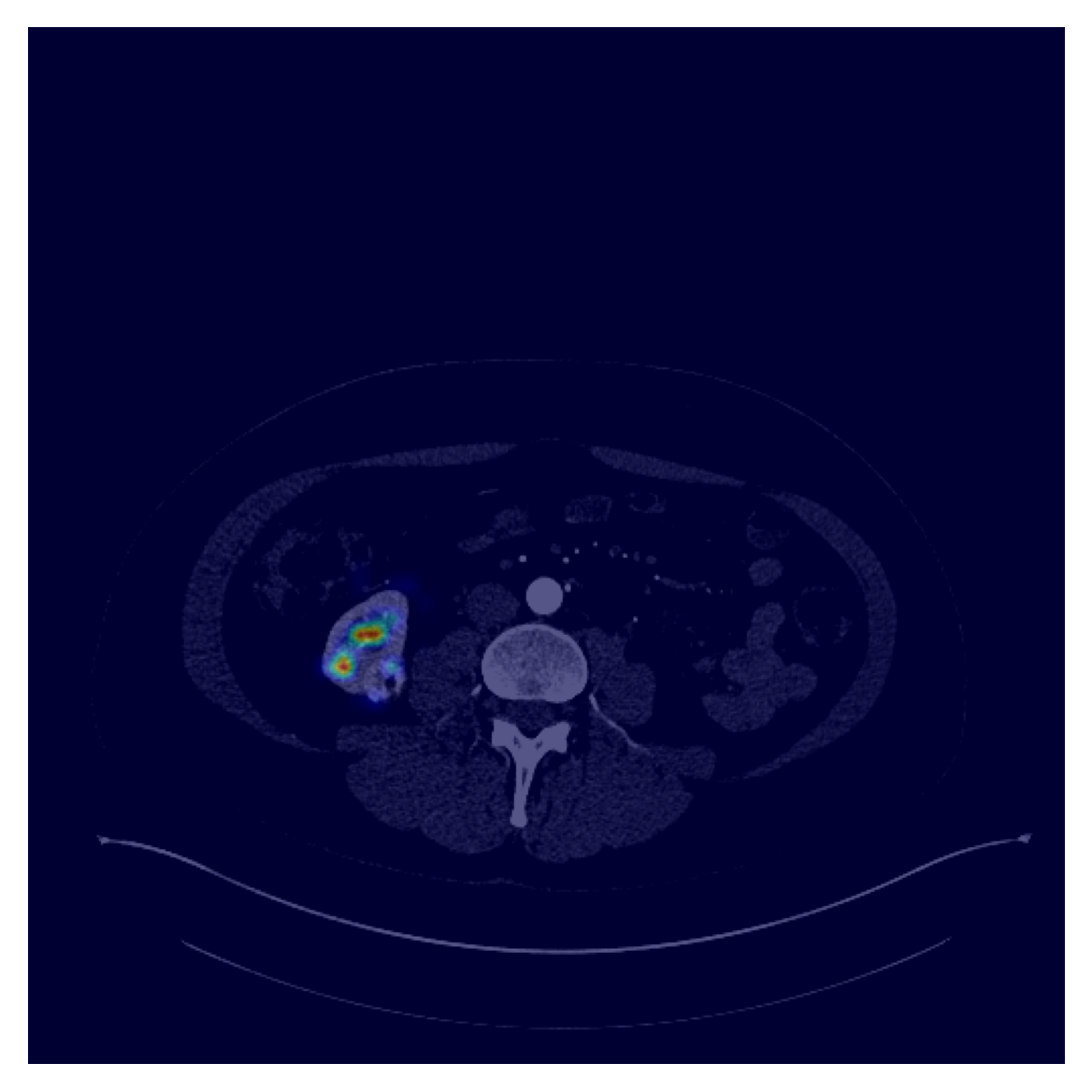}
      \caption{SHRGC}
    \end{subfigure}
    
    \caption{Comparison between \textit{Seg-Grad CAM} \citep{vinogradova_towards_2020} \textbf{(c, g)} and \textit{Seg-HiRes-Grad CAM} \textbf{(d, h)} for a tumor \textbf{(b)} and kidney \textbf{(f)}. The input image comes from the Kits23 dataset \citep{Heller_Isensee_et_al_2020}.}
    \label{fig:cam_vs_hirescam_kits_example}
\end{figure}

\textit{Seg-HiRes-Grad CAM} can highlight pixels that are car-related only. In contrast, \textit{Seg-Grad CAM} highlights coarser regions, including the sky, trees, and street parts. 
Two medical examples demonstrate the difference between both algorithms more distinctly (\cref{fig:cam_vs_hirescam_opg_tooth_18}, \cref{fig:cam_vs_hirescam_kits_example}). In these cases, the baseline method does not produce explainable results, while the result of \textit{Seg-HiRes-Grad CAM} indicates the region around the segmented tooth / tumor accurately. Summarizing these examples, especially segmentations that are close together (such as the teeth) are highlighted more accurately with \textit{Seg-HiRes-Grad CAM} in comparison to \textit{Seg-Grad CAM}. Especially for medical images, the differences of the \glspl{cam} are striking and not as subtle as for natural images. 
Nevertheless, not all segmentations can be explained - even with our proposed \textit{Seg-HiRes-Grad CAM}.

However, the latter is still way more consistent, so it is considered the better visualization method. Exemplary results for, \eg, different pixel sets $\mathcal{M}$, datasets and activation map levels are presented in \cref*{app:appendix}.

\begin{table}
    \caption{Validation-/Test-split results and details of the U-Net for the different datasets (2D slices in case of 3D dataset) we used for our experiments. Hyperparameters are selected empirically, splits are pre-defined or chosen randomly (80, 10, 10), learning rate is set to $3e-3$ and the U-Net depth is four ($512$, $256$, $128$, $64$). The U-Nets are trained for $300$ epochs.}
    \centering 
    \begin{tabular}{|c|c|c|c|c|}
    \toprule
    \thead{\textbf{Dataset}} & \thead{\textbf{F1-Score}} & \thead{\textbf{IoU}} & \thead{\textbf{Resolution}} & \thead{\textbf{Augmentation}} \\ 
    \hline
    \makecell{Cityscapes \\ \citep{cordts_cityscapes_2016}} & \makecell{$0.865$} & \makecell{$0.774$} & \makecell{$512 \times 1024$} & \makecell{None}  \\

    \makecell{OPG \\ \citep{jader_deep_2018}} & \makecell{$0.959$} & \makecell{$$0.921$$} & \makecell{$560 \times 992$} & \makecell{Vertical Mirroring}  \\

    \makecell{Kits23 \\ \citep{Heller_Isensee_et_al_2020}} & \makecell{$0.996$} & \makecell{$0.993$} & \makecell{$512 \times 512$} & \makecell{None}  \\
    \bottomrule
    \end{tabular}
    \label{tab:training_details}
\end{table}

\section{Discussion}
\label{discussion}
Our proposed \textit{Seg-HiRes-Grad CAM} explains salient regions more precisely and accurately in comparison to \textit{Seg-Grad CAM}. Misinterpretations due to the visualization method can thus be better excluded as existing work describes for classification tasks \citep{draelos_use_2021}. Accordingly, \textit{HiRes CAM} can also be applied to segmentation tasks by implementing \textit{Seg-HiRes-Grad CAM}, which provides more transparent results than \textit{Seg-Grad CAM}. Particularly in a medical setting, this difference of explanation can be of strong importance. 
On the other hand, certain limitations have to be addressed. First, the runtime is a general limitation of \gls{cam} algorithms for segmentation tasks. Compared to classification-based \gls{cam} visualization algorithms, the segmentation-based algorithms need more time to produce the heatmap(s) due to the pixel-level probabilities in semantic segmentation tasks instead of a single class distribution for an entire image in the case of classification tasks. 
Second, it should be noted that the success of \gls{cam} algorithms depends on the input image resolution: The method fails for minimal image resolutions in combination with \glspl{cnn}. Small image resolutions are common practice due to \gls{gpu} limitations, especially for 3D data. These cause a minimal resolution of the feature map of the deepest layer of a U-Net. If this feature map is spatially too small, too much detail is lost for an accurate representation of the gradients to be possible. 
Last, the \gls{cam} visualization depends on the segmentation result. Consequently, false negative or false positive segmented pixels will result in less explainable results when \textit{Seg-HiRes-Grad CAM} or similar methods are used.

\section{Conclusion and Future Work}
\label{conclusion}
In this work, we propose a semantic segmentation \gls{cam} visualization method (\textit{Seg-HiRes-Grad CAM}), an extension of \textit{Seg-Grad CAM} by combining it with the classification-based \textit{HiRes CAM}. The proposed method accurately highlights salient regions and delivers more explainable results especially when medical datasets are used. We demonstrate that the on \textit{Grad CAM} based method \textit{Seg-Grad CAM} has the same disadvantage of highlighting misleading regions as Draelos and Carin \citep{draelos_use_2021} elaborated for classification tasks. In contrast, our transfer of \textit{HiRes CAM} \citep{draelos_use_2021} to segmentation tasks generates more consistent results. Also, further transfers to the variety of classification-based methods could be drawn and quantitative methods such as \gls{roar} \citep{Hooker_roar} could be used for enhanced comparisons.

%%%%%%%%%%%%%%%%%%%%%%%%%%%%%%%%%%%%%%%%%%%%%%%%%%%%%%%%%%%%%%%%%%%%%%%
% Mandatory Sections. Please complete, especially for final publication
%%%%%%%%%%%%%%%%%%%%%%%%%%%%%%%%%%%%%%%%%%%%%%%%%%%%%%%%%%%%%%%%%%%%%%%

% Acknowledgements.
% Please include any funding, intellectual contributions not included in the authorship, and any other acknowledgements.
%\acks{This work was supported by grants X, Y and Z. We also acknowledge %important conversations with our colleagues A, B and C.}

% Ethical Standards.
% Please edit with the appropriate ethics considerations for your work. Include any pertinent IRB information, etc.
%
% Please note that the submission requirements included:
% The work presented must follow appropriate ethical standards in conducting research and writing the manuscript, following all applicable laws and regulations regarding treatment of animals or human subjects.
\ethics{The work follows appropriate ethical standards in conducting research and writing the manuscript, following all applicable laws and regulations regarding treatment of animals or human subjects.}

% Conflict of Interest
% Declaration of possible conflicts of interest: Authors must disclose any financial, organisational, commercial or personal conflicts of interest that might bias their work.
% If no conflicts, please say "We declare we don't have conflicts of interest."
%\hl{TODO: Should the similar Paper by Syed Nouman Hasany (Seg-XRes-CAM: Explaining Spatially Local Regions in Image Segmentation) be noted here?} \\ 
\coi{We declare we don't have conflicts of interest.}

% Data availability
\data{All datasets utilized in this study are publicly available and can be accessed freely. However, access to some of these datasets requires prior registration or application for access due to the sensitive nature of the data or to comply with data protection regulations. Despite the public availability of the datasets, direct sharing of the data by the authors is not permissible due to copyright and licensing restrictions imposed by the data providers. We encourage interested researchers to obtain the data directly from the respective repositories, adhering to the specified access procedures and usage policies.}
%\data{Authors submitting articles to \textsc{Melba} are required to include a Data Availability Statement in their manuscripts. The Data Availability Statement should clearly indicate whether the data supporting the findings of the study are available and, if so, how readers can access them. If the data are not available, authors should provide a brief justification for not sharing the data.}

\bibliography{main}

\begin{thebibliography}{25}
\providecommand{\natexlab}[1]{#1}
\providecommand{\url}[1]{\texttt{#1}}
\expandafter\ifx\csname urlstyle\endcsname\relax
  \providecommand{\doi}[1]{doi: #1}\else
  \providecommand{\doi}{doi: \begingroup \urlstyle{rm}\Url}\fi

\bibitem[Agarwal et~al.(2021)Agarwal, Jabbari, Agarwal, Upadhyay, Wu, and Lakkaraju]{agarwal_2021}
Sushant Agarwal, Shahin Jabbari, Chirag Agarwal, Sohini Upadhyay, Steven Wu, and Himabindu Lakkaraju.
\newblock Towards the unification and robustness of perturbation and gradient based explanations.
\newblock In Marina Meila and Tong Zhang, editors, \emph{Proceedings of the 38th International Conference on Machine Learning, {ICML} 2021, 18-24 July 2021, Virtual Event}, volume 139 of \emph{Proceedings of Machine Learning Research}, pages 110--119. {PMLR}, 2021.

\bibitem[Chattopadhyay et~al.(2018)Chattopadhyay, Sarkar, Howlader, and Balasubramanian]{chattopadhay_grad-cam_2018}
Aditya Chattopadhyay, Anirban Sarkar, Prantik Howlader, and Vineeth~N. Balasubramanian.
\newblock Grad-cam++: Generalized gradient-based visual explanations for deep convolutional networks.
\newblock In \emph{2018 {IEEE} Winter Conference on Applications of Computer Vision, {WACV} 2018, Lake Tahoe, NV, USA, March 12-15, 2018}, pages 839--847. {IEEE} Computer Society, 2018.
\newblock \doi{10.1109/WACV.2018.00097}.

\bibitem[Chen et~al.(2022)Chen, Gomez, Huang, and Unberath]{Chen_Gomez_Huang_Unberath_2022}
Haomin Chen, Catalina Gomez, Chien-Ming Huang, and Mathias Unberath.
\newblock Explainable medical imaging ai needs human-centered design: guidelines and evidence from a systematic review.
\newblock \emph{npj Digital Medicine}, 5\penalty0 (1):\penalty0 156, Oct 2022.
\newblock ISSN 2398-6352.
\newblock \doi{10.1038/s41746-022-00699-2}.

\bibitem[Cordts et~al.(2016)Cordts, Omran, Ramos, Rehfeld, Enzweiler, Benenson, Franke, Roth, and Schiele]{cordts_cityscapes_2016}
Marius Cordts, Mohamed Omran, Sebastian Ramos, Timo Rehfeld, Markus Enzweiler, Rodrigo Benenson, Uwe Franke, Stefan Roth, and Bernt Schiele.
\newblock The cityscapes dataset for semantic urban scene understanding.
\newblock In \emph{2016 {IEEE} Conference on Computer Vision and Pattern Recognition, {CVPR} 2016, Las Vegas, NV, USA, June 27-30, 2016}, pages 3213--3223. {IEEE} Computer Society, 2016.
\newblock \doi{10.1109/CVPR.2016.350}.

\bibitem[Desai and Ramaswamy(2020)]{desai_ablation-cam_2020}
Saurabh Desai and Harish~G. Ramaswamy.
\newblock Ablation-cam: Visual explanations for deep convolutional network via gradient-free localization.
\newblock In \emph{{IEEE} Winter Conference on Applications of Computer Vision, {WACV} 2020, Snowmass Village, CO, USA, March 1-5, 2020}, pages 972--980. {IEEE}, 2020.
\newblock \doi{10.1109/WACV45572.2020.9093360}.

\bibitem[Draelos and Carin(2021)]{draelos_use_2021}
Rachel~Lea Draelos and Lawrence Carin.
\newblock Use {HiResCAM} instead of {Grad}-{CAM} for faithful explanations of convolutional neural networks.
\newblock \emph{ArXiv}, November 2021.

\bibitem[Fu et~al.(2020)Fu, Hu, Dong, Guo, Gao, and Li]{fu_axiom-based_2020}
Ruigang Fu, Qingyong Hu, Xiaohu Dong, Yulan Guo, Yinghui Gao, and Biao Li.
\newblock Axiom-based grad-cam: Towards accurate visualization and explanation of cnns.
\newblock In \emph{31st British Machine Vision Conference 2020, {BMVC} 2020, Virtual Event, UK, September 7-10, 2020}. {BMVA} Press, 2020.

\bibitem[Hasany et~al.(2023)Hasany, Petitjean, and Mériaudeau]{Hasany_Petitjean_Meriaudeau_2023}
Syed~Nouman Hasany, Caroline Petitjean, and Fabrice Mériaudeau.
\newblock Seg-xres-cam: Explaining spatially local regions in image segmentation.
\newblock In \emph{Proceedings of the IEEE/CVF Conference on Computer Vision and Pattern Recognition (CVPR) Workshops}, page 3733–3738, June 2023.

\bibitem[Heller et~al.(2020)Heller, Isensee, Maier-Hein, Hou, Xie, Li, Nan, Mu, Lin, Han, et~al.]{Heller_Isensee_et_al_2020}
Nicholas Heller, Fabian Isensee, Klaus~H Maier-Hein, Xiaoshuai Hou, Chunmei Xie, Fengyi Li, Yang Nan, Guangrui Mu, Zhiyong Lin, Miofei Han, et~al.
\newblock The state of the art in kidney and kidney tumor segmentation in contrast-enhanced ct imaging: Results of the kits19 challenge.
\newblock \emph{Medical Image Analysis}, page 101821, 2020.

\bibitem[Hooker et~al.(2019)Hooker, Erhan, Kindermans, and Kim]{Hooker_roar}
Sara Hooker, Dumitru Erhan, Pieter{-}Jan Kindermans, and Been Kim.
\newblock A benchmark for interpretability methods in deep neural networks.
\newblock In Hanna~M. Wallach, Hugo Larochelle, Alina Beygelzimer, Florence d'Alch{\'{e}}{-}Buc, Emily~B. Fox, and Roman Garnett, editors, \emph{Advances in Neural Information Processing Systems 32: Annual Conference on Neural Information Processing Systems 2019, NeurIPS 2019, December 8-14, 2019, Vancouver, BC, Canada}, pages 9734--9745, 2019.

\bibitem[Ivanovs et~al.(2021)Ivanovs, Kadikis, and Ozols]{ivanovs_2021}
Maksims Ivanovs, Roberts Kadikis, and Kaspars Ozols.
\newblock Perturbation-based methods for explaining deep neural networks: {A} survey.
\newblock \emph{Pattern Recognit. Lett.}, 150:\penalty0 228--234, 2021.
\newblock \doi{10.1016/j.patrec.2021.06.030}.

\bibitem[Jader et~al.(2018)Jader, Fontineli, Ruiz, Abdalla, Pithon, and Oliveira]{jader_deep_2018}
Gil Jader, Jefferson Fontineli, Marco Ruiz, Kalyf Abdalla, Matheus Pithon, and Luciano Oliveira.
\newblock Deep instance segmentation of teeth in panoramic x-ray images.
\newblock In \emph{31st {SIBGRAPI} Conference on Graphics, Patterns and Images, {SIBGRAPI} 2018, Paran{\'{a}}, Brazil, October 29 - Nov. 1, 2018}, pages 400--407. {IEEE} Computer Society, 2018.
\newblock \doi{10.1109/SIBGRAPI.2018.00058}.

\bibitem[Jiang et~al.(2021)Jiang, Zhang, Hou, Cheng, and Wei]{jiang_layercam_2021}
Peng{-}Tao Jiang, Chang{-}Bin Zhang, Qibin Hou, Ming{-}Ming Cheng, and Yunchao Wei.
\newblock Layercam: Exploring hierarchical class activation maps for localization.
\newblock \emph{{IEEE} Trans. Image Process.}, 30:\penalty0 5875--5888, 2021.
\newblock \doi{10.1109/TIP.2021.3089943}.

\bibitem[Jung and Oh(2021)]{jung_towards_2021}
Hyungsik Jung and Youngrock Oh.
\newblock Towards better explanations of class activation mapping.
\newblock In \emph{2021 {IEEE/CVF} International Conference on Computer Vision, {ICCV} 2021, Montreal, QC, Canada, October 10-17, 2021}, pages 1316--1324. {IEEE}, 2021.
\newblock \doi{10.1109/ICCV48922.2021.00137}.

\bibitem[Lundberg and Lee(2017)]{lundberg_shap}
Scott~M. Lundberg and Su{-}In Lee.
\newblock A unified approach to interpreting model predictions.
\newblock In Isabelle Guyon, Ulrike von Luxburg, Samy Bengio, Hanna~M. Wallach, Rob Fergus, S.~V.~N. Vishwanathan, and Roman Garnett, editors, \emph{Advances in Neural Information Processing Systems 30: Annual Conference on Neural Information Processing Systems 2017, December 4-9, 2017, Long Beach, CA, {USA}}, pages 4765--4774, 2017.

\bibitem[Muhammad and Yeasin(2020)]{muhammad_eigen-cam_2020}
Mohammed~Bany Muhammad and Mohammed Yeasin.
\newblock Eigen-cam: Class activation map using principal components.
\newblock In \emph{2020 International Joint Conference on Neural Networks, {IJCNN} 2020, Glasgow, United Kingdom, July 19-24, 2020}, pages 1--7. {IEEE}, 2020.
\newblock \doi{10.1109/IJCNN48605.2020.9206626}.

\bibitem[Ronneberger et~al.(2015)Ronneberger, Fischer, and Brox]{ronneberger_u-net_2015}
Olaf Ronneberger, Philipp Fischer, and Thomas Brox.
\newblock U-net: Convolutional networks for biomedical image segmentation.
\newblock In Nassir Navab, Joachim Hornegger, William M.~Wells III, and Alejandro~F. Frangi, editors, \emph{Medical Image Computing and Computer-Assisted Intervention - {MICCAI} 2015 - 18th International Conference Munich, Germany, October 5 - 9, 2015, Proceedings, Part {III}}, volume 9351 of \emph{Lecture Notes in Computer Science}, pages 234--241. Springer, 2015.
\newblock \doi{10.1007/978-3-319-24574-4\_28}.

\bibitem[Selvaraju et~al.(2017)Selvaraju, Cogswell, Das, Vedantam, Parikh, and Batra]{selvaraju_grad-cam_2017}
Ramprasaath~R. Selvaraju, Michael Cogswell, Abhishek Das, Ramakrishna Vedantam, Devi Parikh, and Dhruv Batra.
\newblock Grad-cam: Visual explanations from deep networks via gradient-based localization.
\newblock In \emph{{IEEE} International Conference on Computer Vision, {ICCV} 2017, Venice, Italy, October 22-29, 2017}, pages 618--626. {IEEE} Computer Society, 2017.
\newblock \doi{10.1109/ICCV.2017.74}.

\bibitem[Simonyan et~al.(2014)Simonyan, Vedaldi, and Zisserman]{simonyan_deep_2014}
Karen Simonyan, Andrea Vedaldi, and Andrew Zisserman.
\newblock Deep inside convolutional networks: Visualising image classification models and saliency maps.
\newblock In Yoshua Bengio and Yann LeCun, editors, \emph{2nd International Conference on Learning Representations, {ICLR} 2014, Banff, AB, Canada, April 14-16, 2014, Workshop Track Proceedings}, 2014.

\bibitem[Springenberg et~al.(2015)Springenberg, Dosovitskiy, Brox, and Riedmiller]{springenberg_guided_backprop}
Jost~Tobias Springenberg, Alexey Dosovitskiy, Thomas Brox, and Martin~A. Riedmiller.
\newblock Striving for simplicity: The all convolutional net.
\newblock In Yoshua Bengio and Yann LeCun, editors, \emph{3rd International Conference on Learning Representations, {ICLR} 2015, San Diego, CA, USA, May 7-9, 2015, Workshop Track Proceedings}, 2015.

\bibitem[Srinivas and Fleuret(2019)]{srinivas_full-gradient_2019}
Suraj Srinivas and Fran{\c{c}}ois Fleuret.
\newblock Full-gradient representation for neural network visualization.
\newblock In Hanna~M. Wallach, Hugo Larochelle, Alina Beygelzimer, Florence d'Alch{\'{e}}{-}Buc, Emily~B. Fox, and Roman Garnett, editors, \emph{Advances in Neural Information Processing Systems 32: Annual Conference on Neural Information Processing Systems 2019, NeurIPS 2019, December 8-14, 2019, Vancouver, BC, Canada}, pages 4126--4135, 2019.

\bibitem[Vinogradova et~al.(2020)Vinogradova, Dibrov, and Myers]{vinogradova_towards_2020}
Kira Vinogradova, Alexandr Dibrov, and Gene Myers.
\newblock Towards interpretable semantic segmentation via gradient-weighted class activation mapping (student abstract).
\newblock In \emph{The Thirty-Fourth {AAAI} Conference on Artificial Intelligence, {AAAI} 2020, The Thirty-Second Innovative Applications of Artificial Intelligence Conference, {IAAI} 2020, The Tenth {AAAI} Symposium on Educational Advances in Artificial Intelligence, {EAAI} 2020, New York, NY, USA, February 7-12, 2020}, pages 13943--13944. {AAAI} Press, 2020.

\bibitem[Wang et~al.(2020)Wang, Wang, Du, Yang, Zhang, Ding, Mardziel, and Hu]{wang_score-cam_2020}
Haofan Wang, Zifan Wang, Mengnan Du, Fan Yang, Zijian Zhang, Sirui Ding, Piotr Mardziel, and Xia Hu.
\newblock Score-cam: Score-weighted visual explanations for convolutional neural networks.
\newblock In \emph{2020 {IEEE/CVF} Conference on Computer Vision and Pattern Recognition, {CVPR} Workshops 2020, Seattle, WA, USA, June 14-19, 2020}, pages 111--119. Computer Vision Foundation / {IEEE}, 2020.
\newblock \doi{10.1109/CVPRW50498.2020.00020}.

\bibitem[Zeiler and Fergus(2014)]{zeiler_visualizing_2014}
Matthew~D. Zeiler and Rob Fergus.
\newblock Visualizing and understanding convolutional networks.
\newblock In David~J. Fleet, Tom{\'{a}}s Pajdla, Bernt Schiele, and Tinne Tuytelaars, editors, \emph{Computer Vision - {ECCV} 2014 - 13th European Conference, Zurich, Switzerland, September 6-12, 2014, Proceedings, Part {I}}, volume 8689 of \emph{Lecture Notes in Computer Science}, pages 818--833. Springer, 2014.
\newblock \doi{10.1007/978-3-319-10590-1\_53}.

\bibitem[Zhou et~al.(2016)Zhou, Khosla, Lapedriza, Oliva, and Torralba]{zhou_learning_2016}
Bolei Zhou, Aditya Khosla, {\`{A}}gata Lapedriza, Aude Oliva, and Antonio Torralba.
\newblock Learning deep features for discriminative localization.
\newblock In \emph{2016 {IEEE} Conference on Computer Vision and Pattern Recognition, {CVPR} 2016, Las Vegas, NV, USA, June 27-30, 2016}, pages 2921--2929. {IEEE} Computer Society, 2016.
\newblock \doi{10.1109/CVPR.2016.319}.

\end{thebibliography}

% Manual newpage inserted to improve layout of sample file - not
% needed in general before appendices.
% \newpage

% Appendix is optional
\clearpage
\appendix
%\counterwithin{figure}{section}
\section{Additional Experiments}
\label{app:appendix}

\begin{figure}[ht]
    \centering
    \begin{subfigure}{0.3\textwidth}
      \includegraphics[width=\linewidth]{figures/cityscapes/frankfurt_000000_001236_leftImg8bit.png}
      \caption{Input}
    \end{subfigure}
    \begin{subfigure}{0.3\textwidth}
      \includegraphics[width=\linewidth]{figures/cityscapes/gt_frankfurt_000000_001236_gtFine_labelIds.png}
      \caption{GT}
    \end{subfigure}
    \begin{subfigure}{0.3\textwidth}
      \includegraphics[width=\linewidth]{figures/cityscapes/unet_frankfurt_000000_001236_gtFine_labelIds.png}
      \caption{Pred}
    \end{subfigure}
        
    \begin{subfigure}{0.3\textwidth}
      \includegraphics[width=\linewidth]{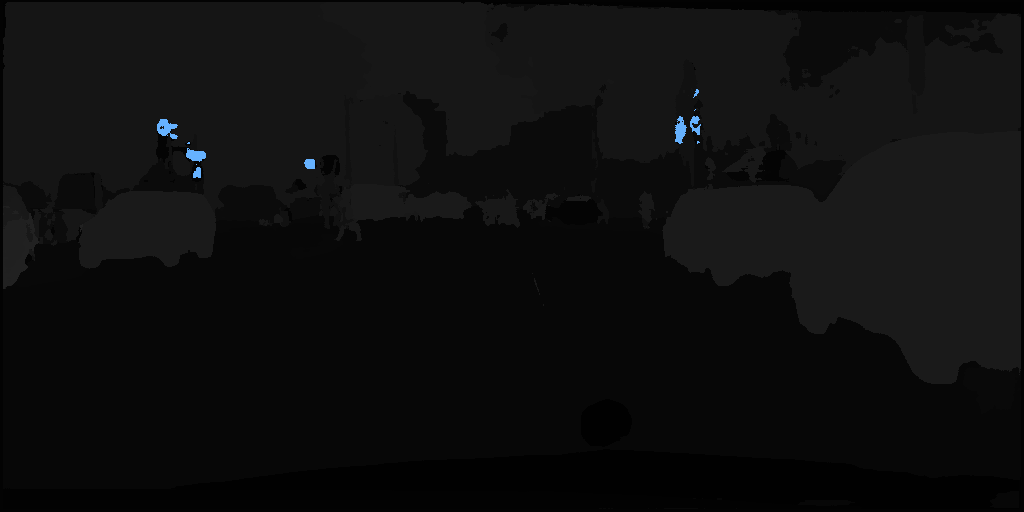}
      \caption{$\mathcal{M}$}
    \end{subfigure}
    \begin{subfigure}{0.3\textwidth}
      \includegraphics[width=\linewidth]{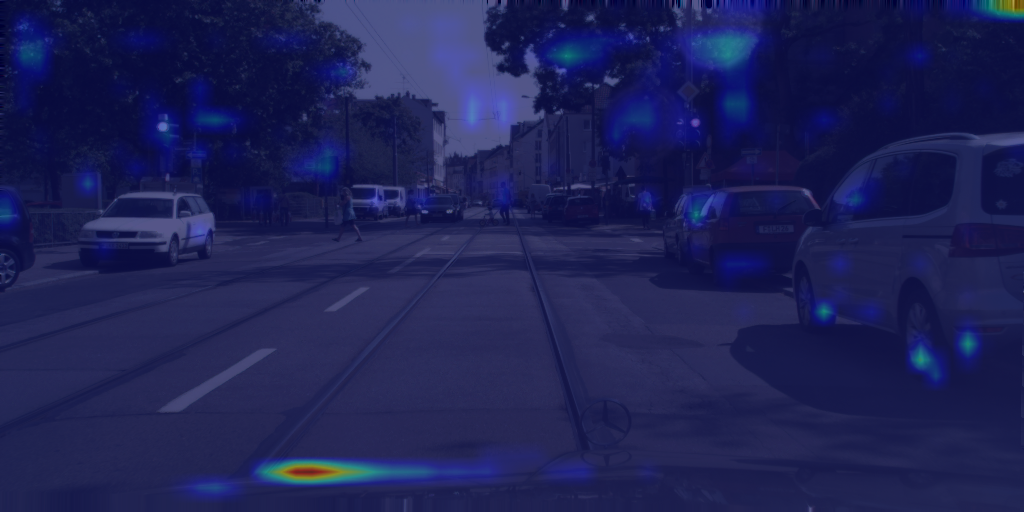}
      \caption{SGC}
    \end{subfigure}
    \begin{subfigure}{0.3\textwidth}
      \includegraphics[width=\linewidth]{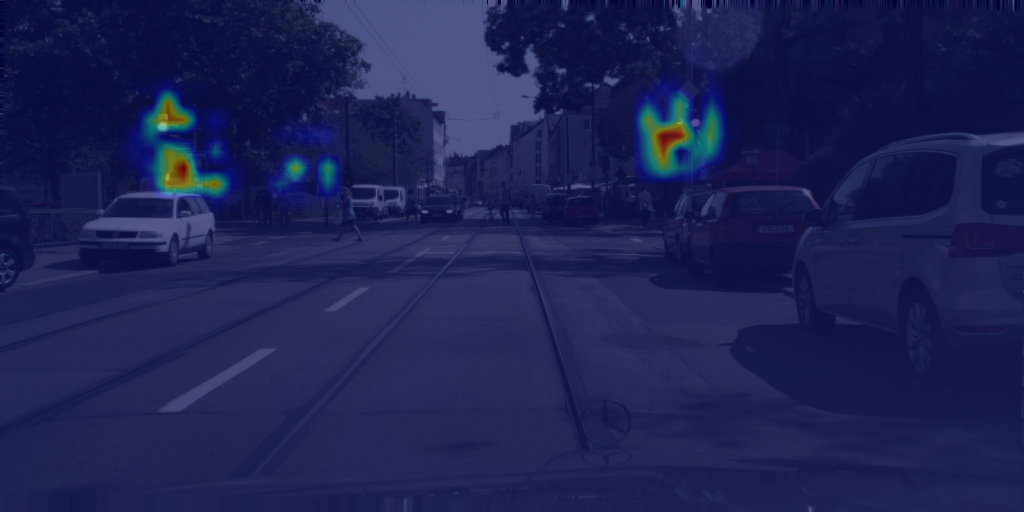}
      \caption{SHRGC}
    \end{subfigure}
    
    \caption{\textit{SGC} \citep{vinogradova_towards_2020} \textbf{(e)} and \textit{SHRGC} \textbf{(f)} for the traffic sign-class \citep{cordts_cityscapes_2016}.}
    \label{fig:cam_vs_hirescam_cityscapes_traffic_sign}
\end{figure}

\begin{figure}[ht]
    \centering
    \begin{subfigure}{0.24\textwidth}
        \captionsetup{justification=centering}

        \caption*{Seg-Grad CAM \citep{vinogradova_towards_2020} ($\mathcal{M}=\text{image}$)}
        
        \includegraphics[width=\linewidth]{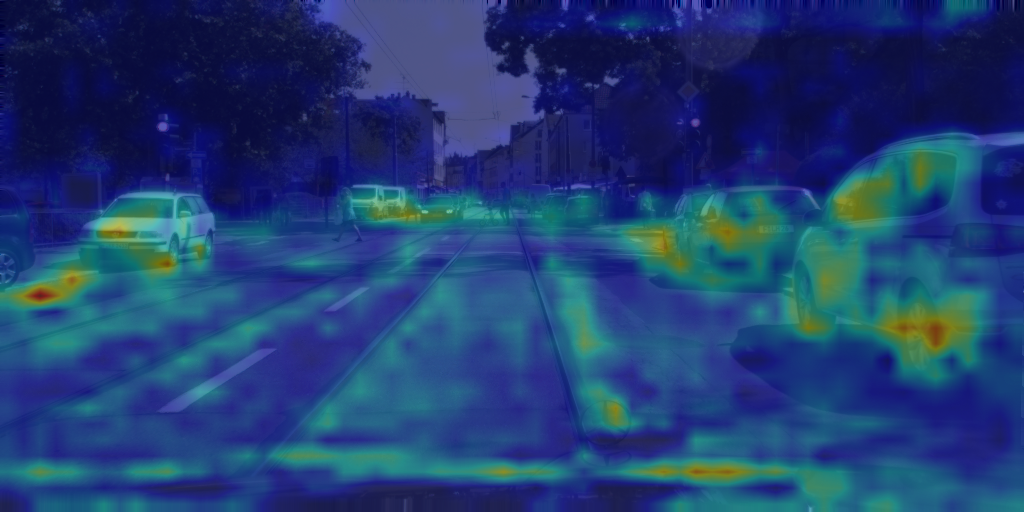}
    \end{subfigure}
    \begin{subfigure}{0.24\textwidth}
        \captionsetup{justification=centering}
        
        \caption*{Seg-Grad CAM \citep{vinogradova_towards_2020}  ($\mathcal{M}=\text{class}$)}
        \includegraphics[width=\linewidth]{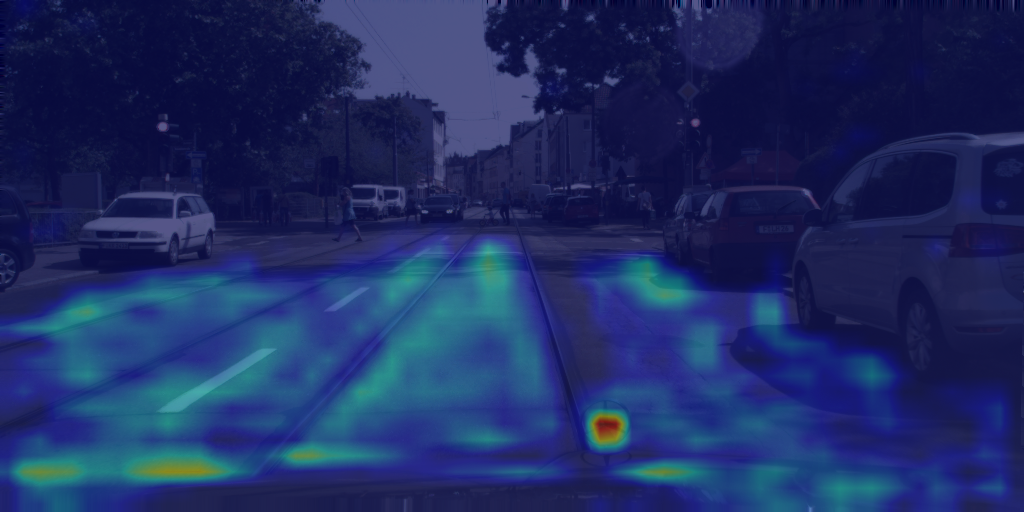}
    \end{subfigure}
    \begin{subfigure}{0.24\textwidth}
        \captionsetup{justification=centering}
        
        \caption*{Seg-HiRes-Grad CAM  ($\mathcal{M}=\text{image}$)}
        \includegraphics[width=\linewidth]{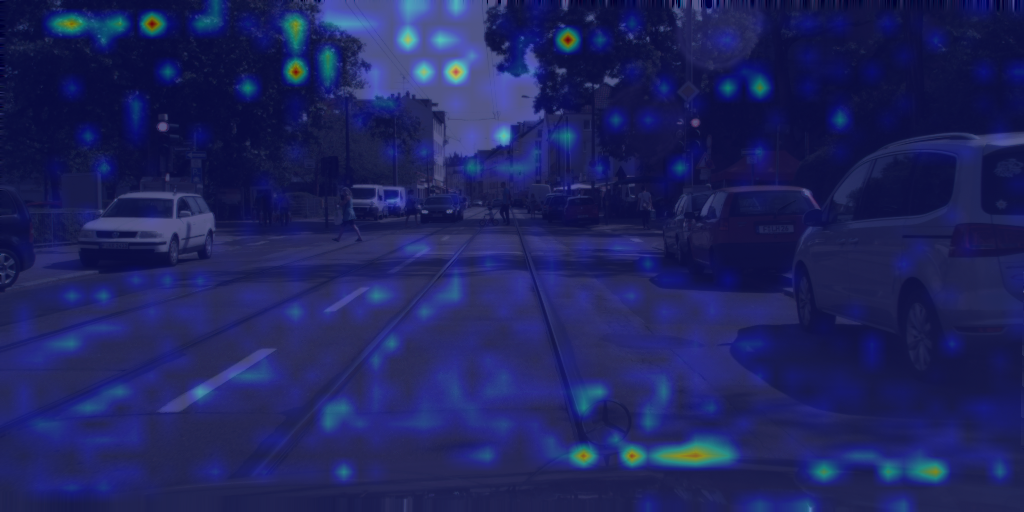}
    \end{subfigure}
    \begin{subfigure}{0.24\textwidth}
        \captionsetup{justification=centering}
        
        \caption*{Seg-HiRes-Grad CAM ($\mathcal{M}=\text{class}$)}
        \includegraphics[width=\linewidth]{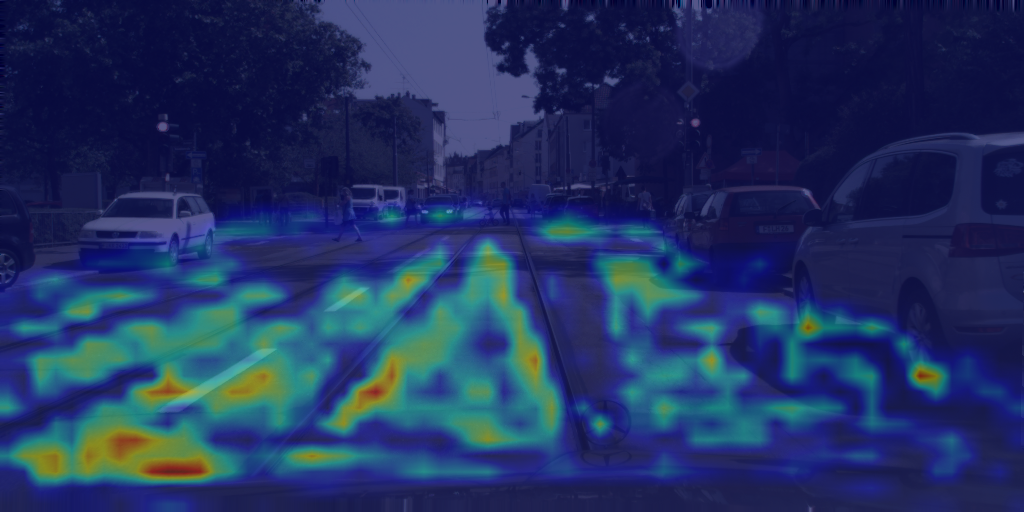}
    \end{subfigure}
    
    \caption{\glspl{cam} for the road-class with different pixel sets and the same ground truth as in \cref{fig:cam_vs_hirescam_cityscapes}.}
    \label{fig:cam_vs_hirescam_cityscapes_road_pixelsets}
\end{figure}

\begin{figure}[ht]
    \centering
    \begin{subfigure}{0.32\textwidth}
      \includegraphics[width=\linewidth]{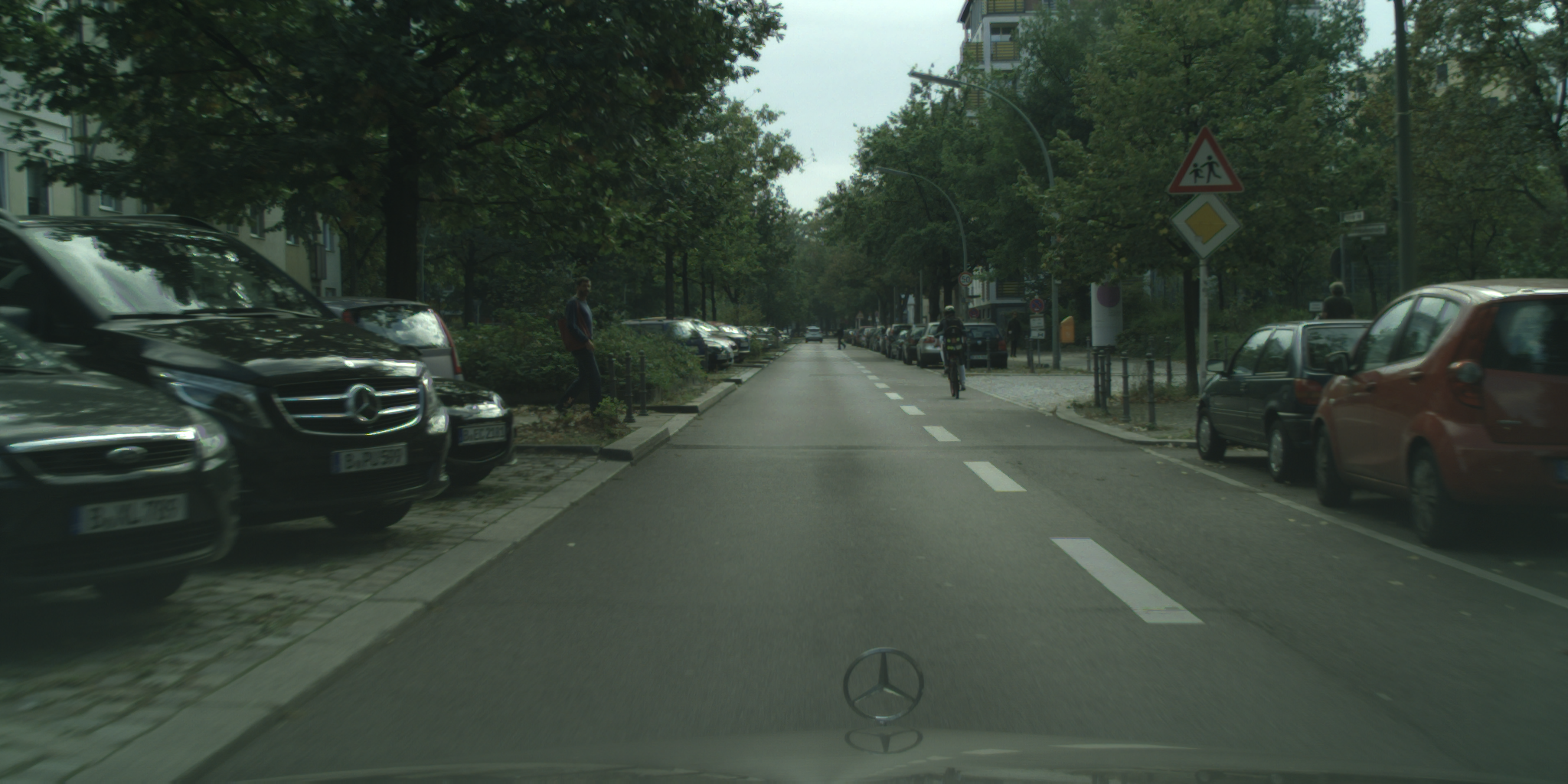}
      \caption{Input image}
    \end{subfigure}
    \begin{subfigure}{0.32\textwidth}
      \includegraphics[width=\linewidth]{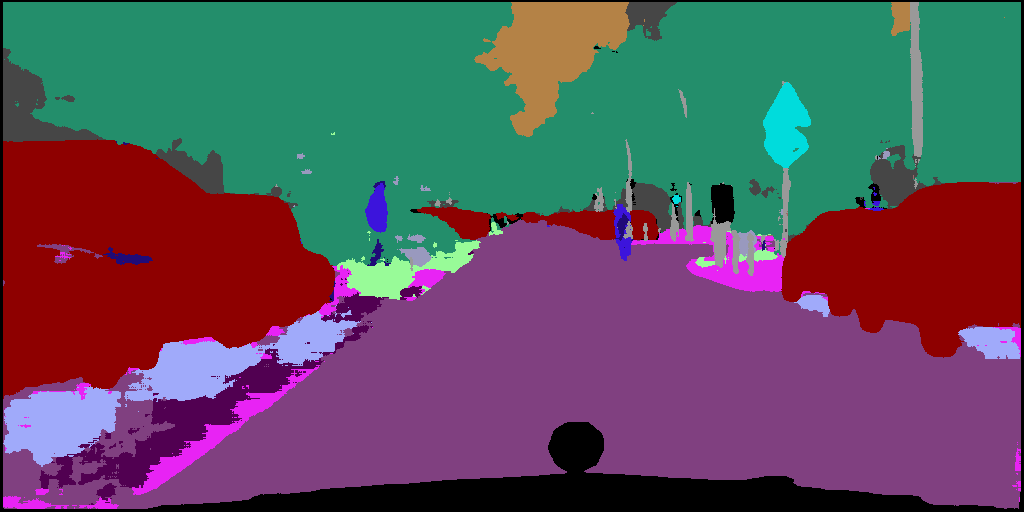}
      \caption{Prediction}
    \end{subfigure}
    \begin{subfigure}{0.32\textwidth}
      \includegraphics[width=\linewidth]{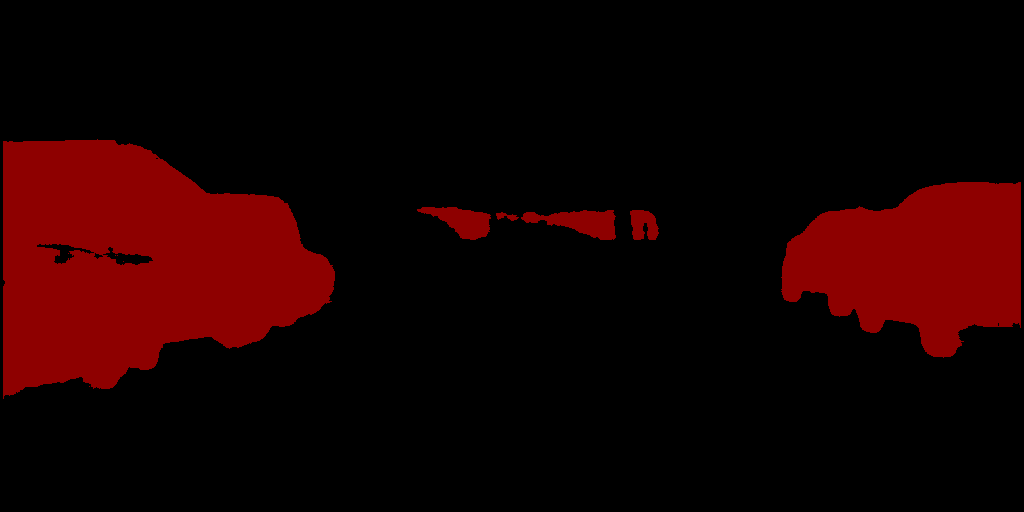}
      \caption{Pixel set $\mathcal{M}$}
    \end{subfigure}
    
    \caption{Input image (\textbf{a}), the predicted segmentation (\textbf{b}) and the pixel set $\mathcal{M}$ for the car-class (\textbf{c}). The ground truth is not available since the Cityscapes \citep{cordts_cityscapes_2016} dataset does not provide the ground truth for the test set.}
    \label{app:input_pixel_set_car}
\end{figure}

\begin{figure}[ht]
    \centering
    \begin{subfigure}{0.24\textwidth}
        \captionsetup{justification=centering}

        \caption*{Seg-Grad CAM \citep{vinogradova_towards_2020} \\ Encoder}
        \includegraphics[width=\linewidth]{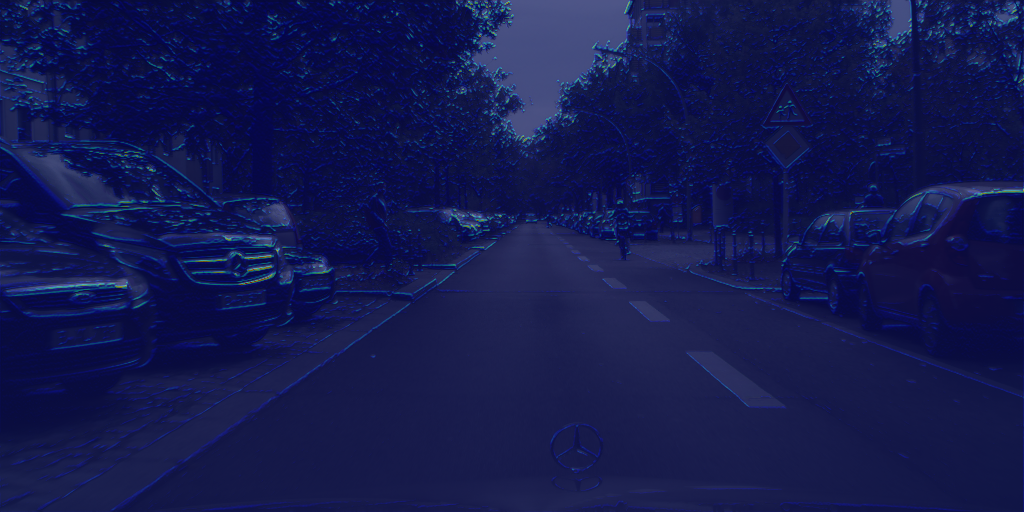}
    \end{subfigure}
    \begin{subfigure}{0.24\textwidth}
        \captionsetup{justification=centering}

        \caption*{Seg-Grad CAM \citep{vinogradova_towards_2020} \\ Decoder}
        \includegraphics[width=\linewidth]{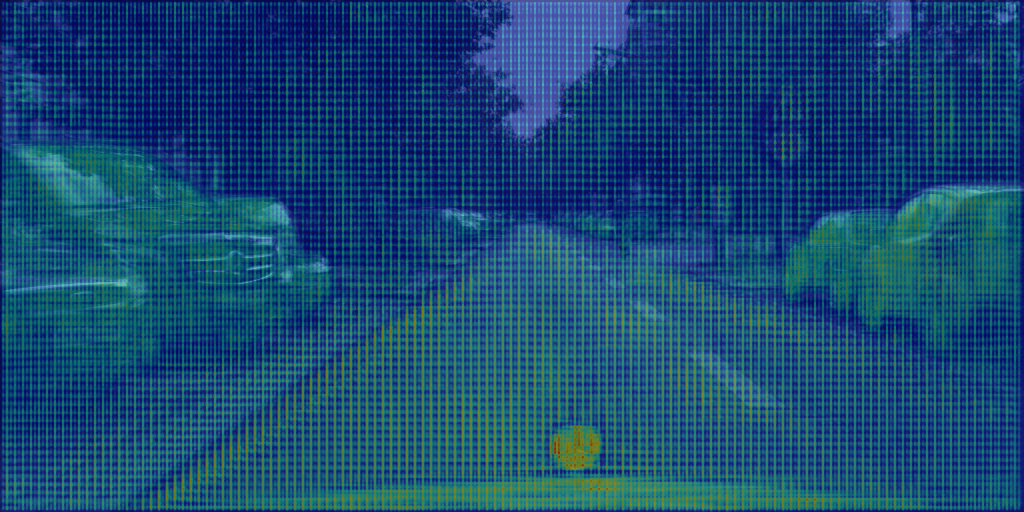}
    \end{subfigure}
    \begin{subfigure}{0.24\textwidth}
        \captionsetup{justification=centering}
        
        \caption*{Seg-HiRes-Grad CAM Encoder}
        \includegraphics[width=\linewidth]{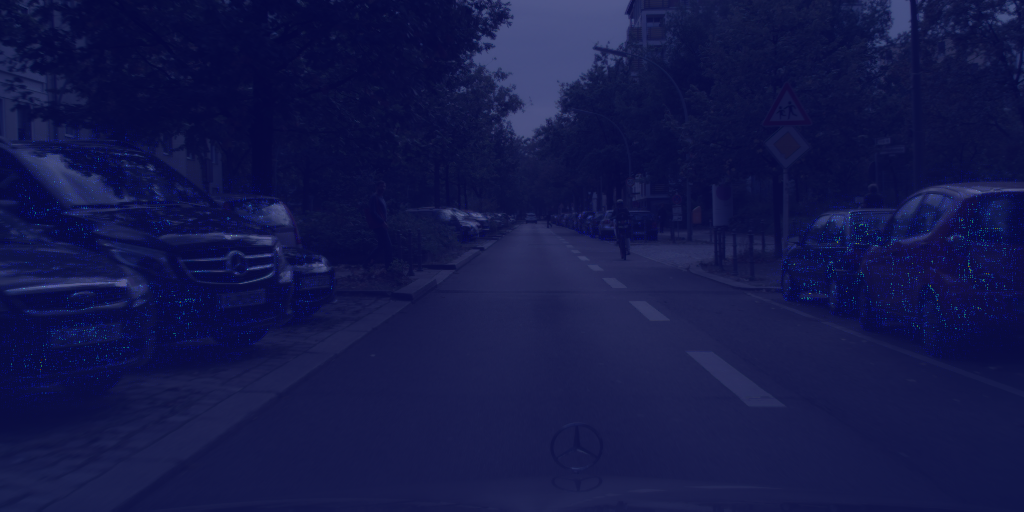}
    \end{subfigure}
    \begin{subfigure}{0.24\textwidth}
        \captionsetup{justification=centering}
        
        \caption*{Seg-HiRes-Grad CAM Decoder}
        \includegraphics[width=\linewidth]{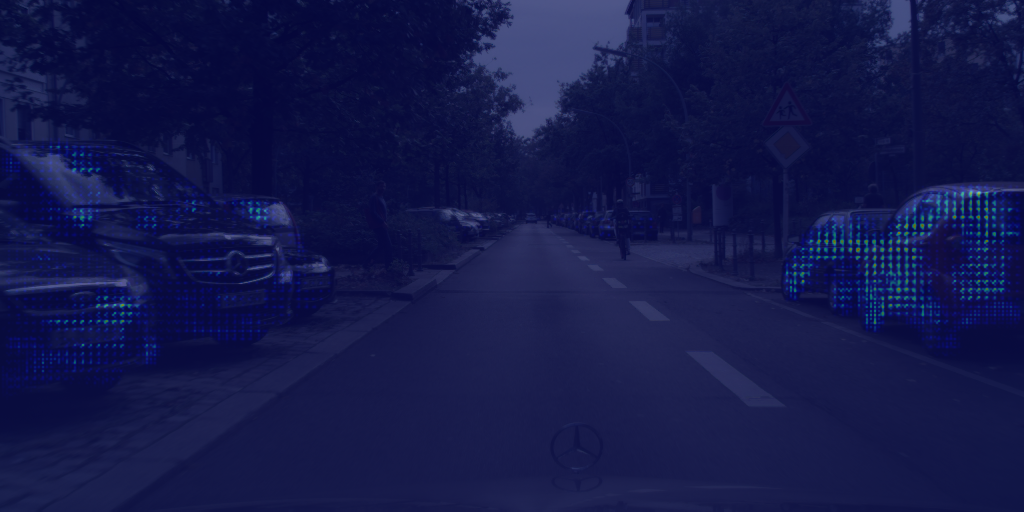}
    \end{subfigure}
    
    \begin{subfigure}{0.24\textwidth}
        \includegraphics[width=\linewidth]{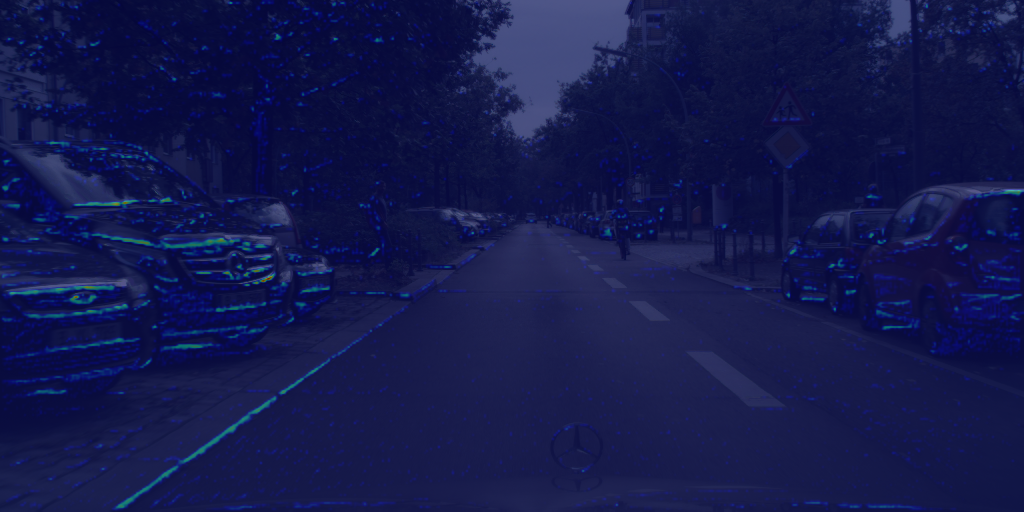}
    \end{subfigure}
    \begin{subfigure}{0.24\textwidth}
        \includegraphics[width=\linewidth]{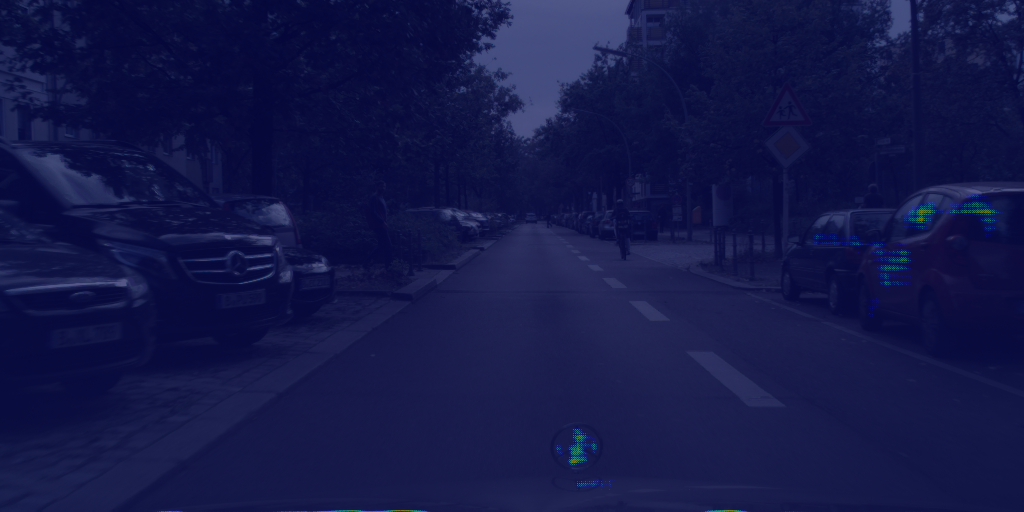}
    \end{subfigure}
    \begin{subfigure}{0.24\textwidth}
        \includegraphics[width=\linewidth]{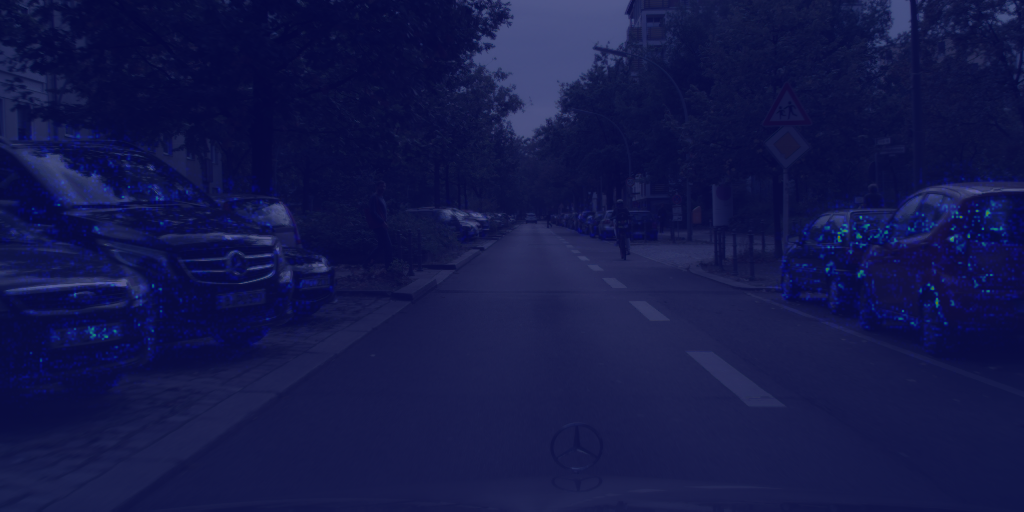}
    \end{subfigure}
    \begin{subfigure}{0.24\textwidth}
        \includegraphics[width=\linewidth]{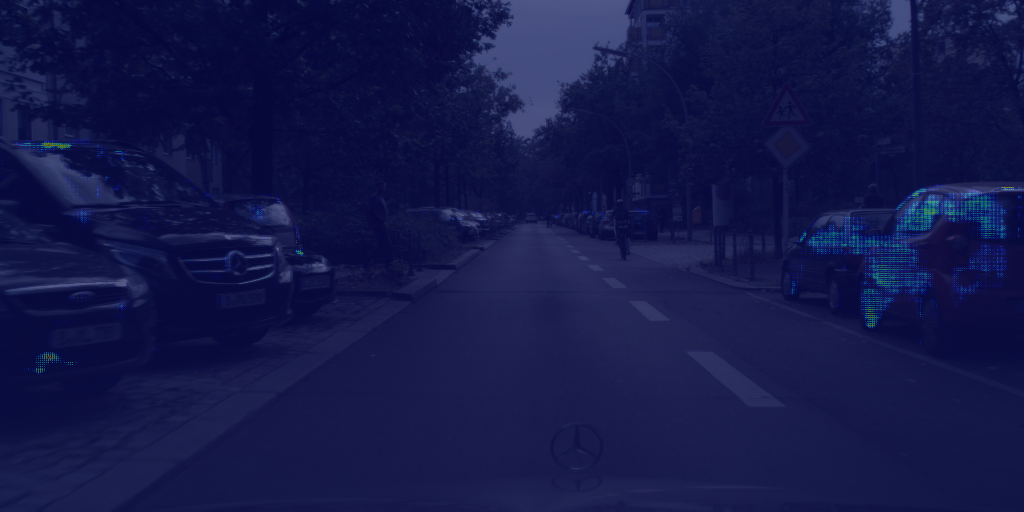}
    \end{subfigure}
        
    \begin{subfigure}{0.24\textwidth}
        \includegraphics[width=\linewidth]{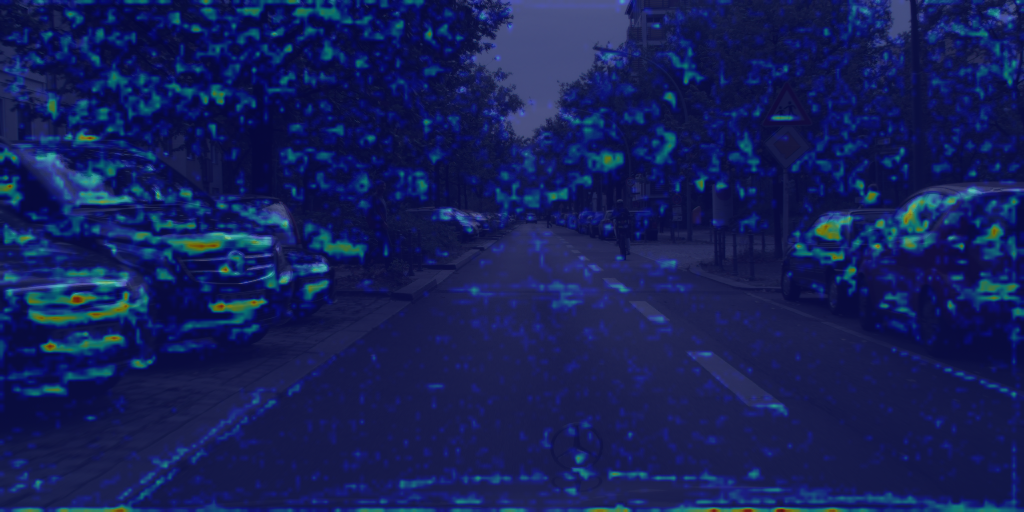}
    \end{subfigure}
    \begin{subfigure}{0.24\textwidth}
        \includegraphics[width=\linewidth]{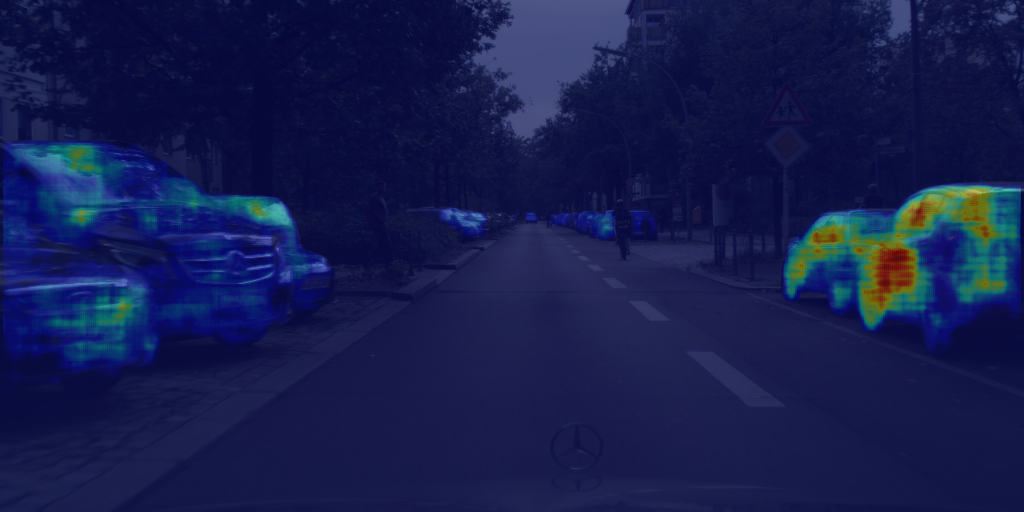}
    \end{subfigure}
    \begin{subfigure}{0.24\textwidth}
        \includegraphics[width=\linewidth]{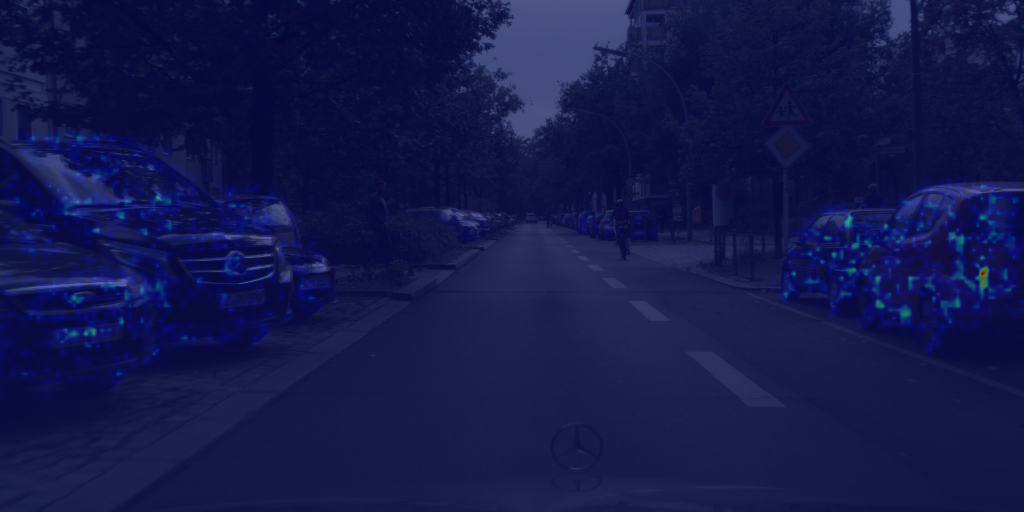}
    \end{subfigure}
    \begin{subfigure}{0.24\textwidth}
        \includegraphics[width=\linewidth]{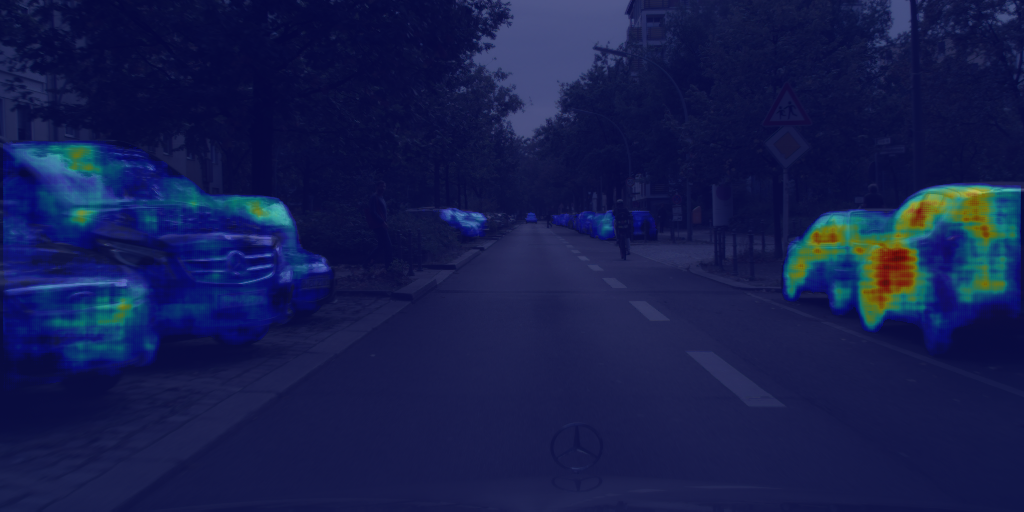}
    \end{subfigure}
    
    \begin{subfigure}{0.24\textwidth}
        \includegraphics[width=\linewidth]{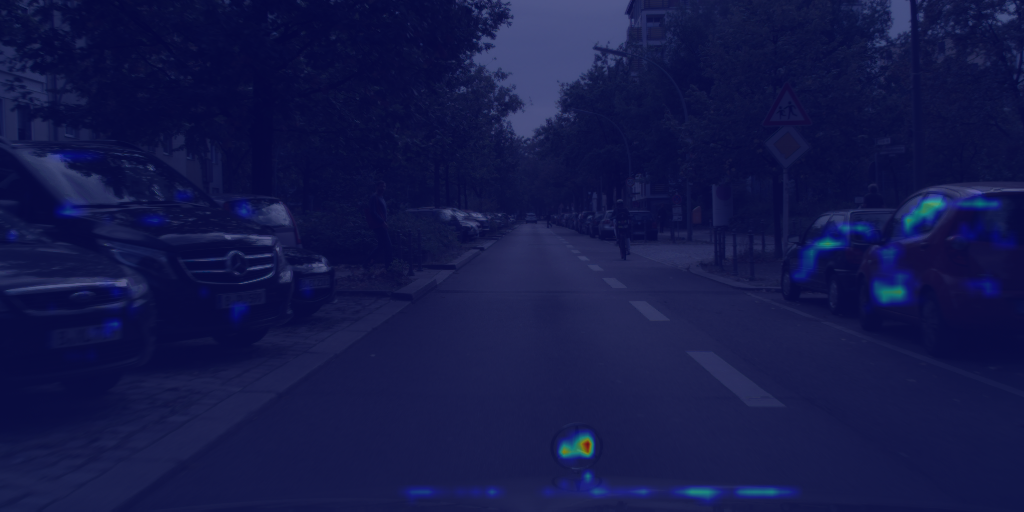}
    \end{subfigure}
    \begin{subfigure}{0.24\textwidth}
        \includegraphics[width=\linewidth]{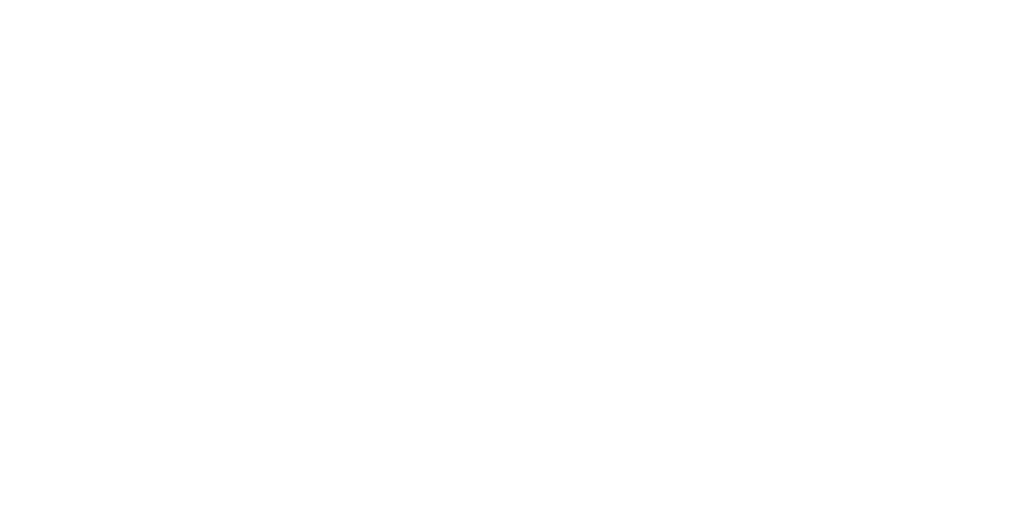}
    \end{subfigure}
    \begin{subfigure}{0.24\textwidth}
        \includegraphics[width=\linewidth]{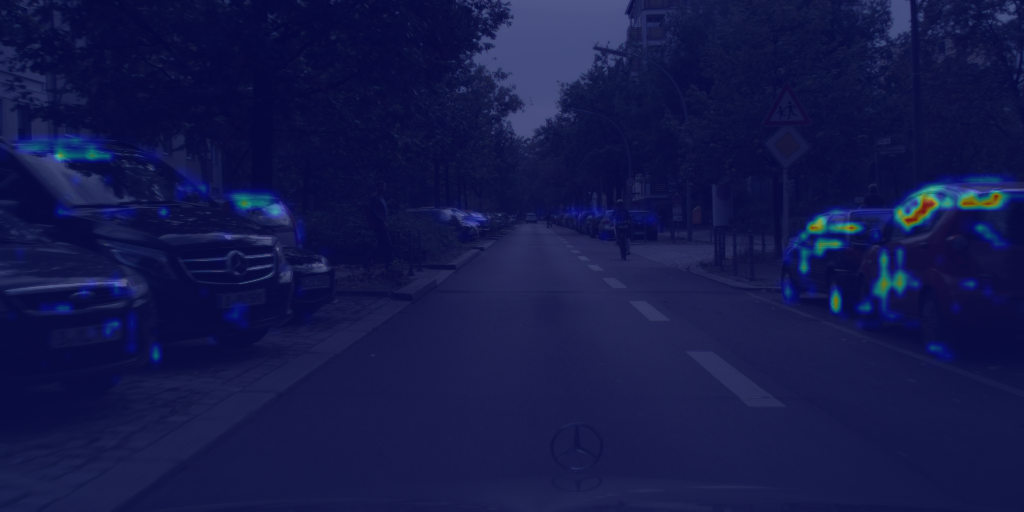}
    \end{subfigure}
    \begin{subfigure}{0.24\textwidth}
        \includegraphics[width=\linewidth]{figures/cityscapes/placeholder.png}
    \end{subfigure}
    
    \caption{\textit{Seg-Grad CAM} \citep{vinogradova_towards_2020} and \textit{Seg-HiRes-Grad CAM} for the image and the pixel set shown in \cref{app:input_pixel_set_car}. The different rows represent the levels of the U-Net \citep{ronneberger_u-net_2015} split into encoder and the decoder. We used the activation maps before the pooling operation but after the convolutional operations of a layer. The last row represents the lowest layer of the U-Net. In this case, the U-Net has a depth of four.}
    \label{app:levels_car_class}
\end{figure}

\end{document}